\documentstyle[12pt]{article}

 \def\be{\begin{equation}} \def\ee{\end{equation}}
 \def\bdm{\begin{displaymath}} \def\edm{\end{displaymath}}
 \def\ba{\begin{array}} \def\ea{\end{array}}
 \def\bea{\begin{eqnarray}} \def\eea{\end{eqnarray}}
 \def\beas{\begin{eqnarray*}} \def\eeas{\end{eqnarray*}}
 \def\disp{\displaystyle}
 \newcommand{\dfrac}[2]{{\displaystyle \frac{#1}{#2}}}
 \newcommand{\tfrac}[2]{{\textstyle \frac{#1}{#2}}}

\makeatletter
\def\section{\@startsection {section}{1}{\z@}{-1.5ex plus -.5ex         
minus -.2ex}{1ex plus .2ex}{\large\bf}}                                 
\makeatother

 \newcommand{\mysection}[1]{\section{#1}\setcounter{equation}{0}}

 \newcounter{remno}

      \def\gr{general relativity}
      
      \def\qg{quantum gravity}

      \def\wrt{with\ respect\ to}

      \def\iff{if\ and\ only\ if}
      \def\ps{phase space}

      \def\bra#1{\langle\, #1\, |}
      \def\d{\partial}
      \def\half{{\textstyle{1\over2}}}
      
      \def\IP#1#2{\langle\, #1\, |\, #2\, \rangle}
      \def\ket#1{|\, #1\, \rangle}
      
      \def\lint{\int\nolimits}
      
      \def\ovr{\overline}
      \def\pb#1{\rlap{\lower1ex\hbox{$\leftarrow$}}#1{}}
      \def\real{{\rm I\!R}}
      \def\rd{{\rm d}}
      
      \def\tw{\widetilde}
      
      \def\wh{\widehat}        
      \newcommand{\fr}[2]{\frac{#1}{#2}}

 \def\Gbar{\hbox{$\ovr{\Gamma}$}}

 \def\cC{\hbox{${\cal C}$}}                    
 
 \def\cV{\hbox{${\cal V}$}}                    
 \def\cVp{\hbox{${\cal V}_{phy}$}}             
 \def\cA{\hbox{$\cal A$}}                      
 \def\cS{\hbox{$\cal S$}}                 
 \def\cAp{\hbox{${\cal A}_{phy}$}}             
 \def\cAs{\hbox{${\cal A}^{(\star)}$}}         
 \def\cAps{\hbox{${\cal A}_{phy}^{(\star)}$}}  

\def\cs{\hbox{$\cal S$}}

\def\Dqp{Dirac quantization procedure}
 \def\rmfrac{{\rm frac}}
\def\ha{\hat{a}} \def\hc{\hat{c}} \def\hJ{\hat{J}} \def\hN{\hat{N}}
\def\hU{\hat{U}}

  \def\OUP{Oxford University Press}
 \def\SV{Springer-\negthinspace Verlag}
 \def\WSS{World Scientific, Singapore}
 \def\pc{{\it private communication}}

 \def\CMP{{Commun.\ Math.\ Phys.\ }}
 \def\CQG{{Class.\ Quantum Grav.\ }}
 
 \def\IJMP{{Int.\ J.\ Mod.\ Phys.\ }}
 \def\IJMPD{{\IJMP D}}
 \def\JMP{{J.\ Math.\ Phys.\ }}
 \def\NP{{Nucl.\ Phys.\ }}
 
 \def\PR{{Phys.\ Rev.\ }}
 \def\PRD{{\PR D}}
 \def\PRL{{\PR Lett.\ }}
 \def\RPP{{Rep.\ Prog.\ Phys.\ }}

 \newcommand{\skhatA}{\hbox{$\skew6\hat{A}$}}
 \newcommand{\skhatf}{\hbox{$\skew5\hat{f}$}}
 \newcommand{\hq}{\hat{q}}
 \newcommand{\hp}{\hat{p}}
 \newcommand{\hz}{\hat{z}}

 \def\Cbar{\bar{\cal C}}    
 \def\tcbar{\tw{\cal C}}    

   \def\phihat{\hat{\phi}}
  \def\ptthat{\hat{p}_\theta} \def\pphihat{\hat{p}_\phi}

 \def\eip{\,\exp(+\fr{i}\hbar \sqrt{R}\,\theta)}
 \def\eim{\,\exp(-\fr{i}\hbar \sqrt{R}\,\theta)}

 \def\Pphihat{\wh{\rm P}_\phi}      
 \def\Pphihatp{\wh{\rm P}_\phi^+}
 \def\Pphihatm{\wh{\rm P}_\phi^-}

 \def\kop{k_1^+} \def\kom{k_1^-}
 \def\ktp{k_2^+} \def\ktm{k_2^-}
 \def\kopb{\ovr{\kop}} \def\komb{\ovr{\kom}}

 \def\Pphi{{\rm P}_\phi}

 \def\tmu{{\tw\mu}}

 \def\bcV{\bar{\cal V}}         
 \def\tcV{\tw{\cal V}}          

\def\rmfrac{{\rm frac}}

\def\ha{\hat{a}} \def\hc{\hat{c}} \def\hJ{\skew4\hat{J}} \def\hN{\hat{N}}
\def\hC{\hat{C}}  \def\hL{\hat{L}}

\def\Sch{Schr\"odinger}

\def\hC{\hat{C}}
\def\hH{\hat{H}}
\def\hU{\hat{U}}
\def\hP{\hat{P}}
\def\hQ{\hat{Q}}

\def\eiq{e^{-\frac{i}\hbar\hH q^0}}

\def\hZ{\hat{Z}}

\begin{document}
\hphantom{a}
\vspace{.25in}

\begin{flushright}
gr-qc/9405073 \\
CGPG-94/6-1
\end{flushright}
\vspace{.5in}

\centerline{\LARGE An algebraic extension of Dirac quantization:}

\centerline{\LARGE Examples}
\vspace{.25in}

\centerline{\large Abhay Ashtekar$^*$ and Ranjeet S. Tate$^{*\dagger}$}
\vspace{.25in}

$*$ Center for Gravitational Physics and Geometry, Physics
Department, Pennsylvania State University, University Park, PA
16802-6300.

$\dagger$ Department of Physics, University of California, Santa
Barbara, CA 93106-9530.

e-mail: ashtekar@phys.psu.edu, rstate@cosmic.physics.ucsb.edu
\vspace{.25in}


\begin{abstract}
An extension of the Dirac procedure for the quantization of
constrained systems is necessary to address certain issues that are
left open in Dirac's original proposal. These issues play an important
role especially in the context of non-linear, diffeomorphism invariant
theories such as general relativity. Recently, an extension of the
required type was proposed by one of us using algebraic quantization
methods. In this paper, the key conceptual and technical aspects of
the algebraic program are illustrated through a number of finite
dimensional examples. The choice of examples and some of the analysis
is motivated by certain peculiar problems endemic to \qg.  However,
prior knowledge of general relativity is {\it not} assumed in the main
discussion. Indeed, the methods introduced and conclusions arrived at
are applicable to any system with first class constraints. In
particular, they resolve certain technical issues which are present
also in the reduced phase space approach to quantization of these
systems.
\end{abstract}

\mysection{Introduction}

A number of systems with first class constraints arise naturally in
mathematical physics: interesting examples are provided by Yang-Mills
theories, general relativity, string theory and topological field
theories. The problem of non-perturbative quantization of such
theories is of considerable interest from both physical and
mathematical perspectives. In the case of QCD for example, one hopes
that a non-perturbative quantization would provide a satisfactory
explanation of the physical phenomenon of confinement. Similarly, a
full quantization of topological field theories is expected to provide
a wealth of information on the topology of low dimensional manifolds
as well as a classification of knots.

There are two major avenues to the non-perturbative quantization of
such systems. The first is the Dirac approach, in which, roughly
speaking, one first ignores the constraints, quantizes the system and
then selects the admissible physical states by demanding that they be
annihilated by the quantum constraint operators. The second is the
reduced phase space method where one first eliminates the constraints
classically and then quantizes the resulting unconstrained system. In
this paper, we shall focus on the Dirac method although some of the
main points we raise and the conclusions we reach are
relevant also to the reduced phase space method.

In an algebraic version of the Dirac quantization procedure \cite{pamd}
--essentially the one followed originally by Dirac-- one first ignores
the constraints and constructs a ``kinematical'' operator algebra
\cA\ starting from the phase space of the system. One then seeks a
representation of \cA\ by operators acting on a complex vector space
\cV. Next, one represents the constraints as concrete operators on the
chosen vector space. Imposition of constraints is then carried out by
finding the kernel of the constraint operators. Since the operators
are all linear, the kernel \cVp\ has a natural vector space structure.
This is the space of {\em physical} states and elements of
\cA\ which map \cV\ to itself are the {\em physical} operators.  In
simple physical systems, one can carry out this program completely.
For source-free Maxwell fields in Minkowski space, for example, the
algebra \cA\ is generated by $\hat{A}(g):=\int d^3x\hat{A}_a(x)g^a(x)$
and $\hat{E}(f):= \int\ d^3x\hat{E}^a(x) f_a(x)$, the smeared vector
potentials and electric fields. For the vector space \cV, one can
choose the space of functionals of vector potentials $\Psi(A)$,
represent $\hat{A}(g)$ by a multiplication operator and $\hat{E}(f)$
by a (directional) derivative (in the direction of $f_a$).  One then
imposes the Gauss constraint: $\int d^3x\ \hat{E}^a
\partial_a \Lambda\ \circ \Psi(A) = 0$ for all suitably regular
functions $\Lambda(x)$.  This equation can be readily solved. It tells
us that $\Psi(A)$ must be gauge invariant. Thus, as one might expect,
the space \cVp\ of physical states is precisely the space of gauge
invariant functions of vector potentials.

This general procedure, however, is incomplete. The most significant
limitation is that it leaves open the issue of finding the inner
product on the space \cVp\ of physical states; no prescription what so
ever is provided. In practice, one generally appeals to suitable
symmetries and asks that the inner product be such that these
symmetries are unitarily implemented in the quantum theory. In the
above example of free Maxwell fields, the required symmetries are
provided by the Poincar\'e group acting on the underlying Minkowski
space-time. However, an appropriate symmetry group is not always
available. In the case of general relativity, for example, there is no
background space-time and hence no analog of the Poincar\'e group%
\footnote{One might seek symmetries directly on the infinite
 dimensional phase space. However, it is known that appropriate
 symmetries fail to exist. For 4-dimensional general relativity, see
 \cite{kk:sym,torre:sym}. The results of \cite{kk:sym} also hold in
 3-dimensional relativity\cite{barbero}.}.
One might imagine using the entire diffeomorphism group. However,
just as the quantum constraint of Maxwell theory requires that all
physical states be gauge invariant, those of general relativity
require that they {\em all} be diffeomorphism invariant. The
diffeomorphism group thus has a trivial action on \cVp; it is unitary
with respect to {\em any} inner product! Thus, the fact that there is
no prescription available to find the inner product is a severe
limitation, especially in the context of diffeomorphism invariant
theories.

A second limitation arises already in the first step, in the
construction of the algebra \cA. In simple examples, such as the free
Maxwell fields discussed above, the phase space is linear and the
construction of \cA\ is straightforward. However, in general, the \ps\
is a genuine manifold, i.e., it does not admit a global chart. In the
classical theory, then, it is {\em not possible} to specify a complete set
of globally defined configuration and momentum variables unless they
are {\em over}complete. We can associate operators with all these
variables. However, then we have to face the overcompleteness squarely
in the quantum theory.  (This situation arises, for example, in
lattice gauge theories, where the Wilson loop variables are
overcomplete.) In the original Dirac program, this issue was not
addressed.

These --and related-- issues were considered by Isham in
\cite{isham:lh}, where he developed a group-theoretic approach to
quantization to resolve these difficulties. In this paper, we will
focus on another approach, based on algebraic methods. \cite{newbook}.
These methods are more general in the sense that, if an appropriate
``canonical group'' on the classical phase space can be found to
complete the Isham program for a given system, key steps of the
algebraic program would be automatically completed. On the other hand,
there also exist other methods to complete the program. Being more
general, however, the algebraic program is correspondingly less
specific: While the group theoretic methods are ``tight'' in the sense
that once the appropriate group is found, very little new input is
needed, as we will see in section 2, the algebraic program needs
inputs at several stages.

In the algebraic approach, the problem of finding the inner product
was addressed through the following prescription: roughly, it states
that the inner product should be such that a complete set of real
classical observables should be represented on $\cVp$ by self-adjoint
operators. It is now known \cite{rendall} that such an inner product,
if it exists, is unique (modulo a multiplicative, overall constant).
The prescription seems rather trivial at first sight. However, it is
both powerful and --in some respects-- subtle. In the case of
3-dimensional general relativity, for example, although there are
again no symmetries \cite{barbero} (and, for a general spatial
topology, no ``deparametrization'' is available), this principle does
select the physical inner-product uniquely. The subtleties can be seen
in the proof of uniqueness \cite{rendall}, as well as in the
occasional occurrence of unforeseen superselected sectors, examples of
which are discussed later in this paper. The second issue, that of
overcompleteness, was addressed by providing a detailed prescription
for the construction of the algebra \cA; the new inputs required to
pass from the classical to the quantum theory were isolated and a
prescription was given to handle the possible overcompleteness. Note
that these two issues arise also in the reduced phase space method:
one again needs a guideline to select the inner product on the space
of physical quantum states, and, since the reduced phase spaces are
typically genuine manifolds without a natural cotangent bundle
structure, one must deal with overcompleteness. Therefore, the
algebraic program developed in \cite{newbook} is relevant to the reduced
phase space approach as well.

The purpose of this paper is to illustrate various aspects of the program
through five examples. The examples are motivated primarily by the
problem of quantum general relativity and each of them mimics one or
more ``peculiarities'' of general relativity. However, for the main
discussion, no prior knowledge of general relativity is required.

In section 2, we summarize the algebraic program of \cite{newbook} on
which the rest of the paper is based, with special attention to the
new ingredients. In section 3, we consider an unconstrained system to
illustrate how the prescription for finding the inner product can
enable one to quantize systems with ``hybrid'' canonical variables, in
which the configuration variable, for example, may be complex but its
conjugate momentum, real. (Such variables arise in 4-dimensional
general relativity and simplify the structure of Einstein's equations
considerably). In sections 4-6, we discuss four constrained systems.
In each case, we carry out the quantization program in detail to
illustrate its various subtle aspects. The example discussed in
section 5 is especially interesting as it illustrates how the program
can be used to resolve ambiguities that may have important physical
consequences. In all these cases, we are motivated primarily by issues
in mathematical physics. Therefore, issues of detailed physical
interpretation and those associated with the measurement theory will
be left untouched.

The program of \cite{newbook} was distilled and refined from a number
of examples. Because of this and because we have tried to make our
discussion here self-contained, there is some inevitable overlap
between \cite{newbook} and the material covered in this paper. The
previous discussion was, however, incomplete in certain respects and
also contained a minor error. We have taken this opportunity to
rectify the situation.

\mysection{Quantization Program}

In this section we will outline the extension of the \Dqp\ for
constrained systems --which constitutes the basis for the rest of the
paper-- and discuss in some detail its key features.  The extension is
based on the algebraic approach to quantum mechanics
\cite{aa:rg,aa:cmp}.

\subsection{The main steps}
Consider a classical system for which the phase space $\Gamma$ is a
real symplectic manifold (which will be finite dimensional in all
examples discussed in the paper). We are particularly interested in
systems which are subject to first class constraints. To quantize such
a system, we proceed in the following steps:

\begin{enumerate}
\item Select a subspace \cS\ of the vector space of all smooth,
complex-valued functions on $\Gamma$ subject to the following
conditions:
 \begin{enumerate}
 \item ${\cal S}$ should be large enough so that any sufficiently regular
function on the phase space can be obtained as (possibly a suitable
limit of) a sum of products of elements $F^{(i)}$ in \cS.
Technically, {\em \cS\ is (locally) {\em complete} if and only if the
gradients of the functions $F$ in \cS\ span the cotangent space of\
$\Gamma$ at each point}. The unit function ``1'' should also be
included in \cS.
 \item ${\cal S}$ should be {\em closed under Poisson brackets}, i.e.\ for
all functions $F,G$ in \cS, their Poisson bracket $\{ F,G\}$ should
also be an element of \cS.
 \item Finally, \cS\ should be {\em closed under complex conjugation},
i.e.\  for all $F$ in \cS, the complex conjugate $\bar{F}$ should be a
function in \cS.
 \end{enumerate}
Each function in \cs\ is to be regarded as an {\em elementary
classical variable} which is to have an {\em unambiguous} quantum
analog \cite{aa:cmp}. The first requirement on \cS\ ensures that the
space of the resulting elementary quantum operators is ``large
enough,'' the second enables us to define commutators between these
operators unambiguously while the third will lead us to the
$\star$-relations between these operators. It is often the case that
in order to satisfy the completeness requirement, one is forced (say
by the non-trivial topology of the \ps) to include {\em more}
functions in the set \cS\ than the dimension of the \ps. In this case,
there are algebraic relations between them.
\item Associate with each element $F$ in \cS\ an abstract operator
$\hat{F}$. Construct the free algebra generated by these {\em
elementary quantum operators}. Impose on it the canonical commutation
relations, $[\hat{F}, \hat{G}] = i\hbar\widehat{\{F,G\}}$, and, if
necessary, also the (anti-commutation) relations which capture the
algebraic identities satisfied by the elementary classical variables.
(The details are discussed in section 2.2.) Denote the resulting algebra by
\cA.
\item On this algebra, introduce an involution operation%
\footnote{An involution operation on \cA\ is a map $\star$ from \cA\ to
 itself satisfying the following three conditions for all $A$ and $B$
 in \cA: i) $(A +\lambda B)^\star = A^\star + \bar\lambda B^\star$,
 where $\lambda$ is any complex number and $\bar\lambda$ its complex
 conjugate; ii)$(AB)^\star = B^\star A^\star$; and, iii)
 $(A^\star)^\star = A$.},
denoted by $\star$, by requiring that if two
elementary classical variables $F$ and $G$ are related by $\bar{F} =
G$ (where $\bar{F}$ denotes the complex conjugate of $F$), then
$\hat{F}{}^\star=\hat{G}$ in \cA.  Denote the resulting
$\star$-algebra by \cAs. As can be seen from its definition, the
$\star$-relation on \cAs\ reflects the reality relation between the
elementary variables.
\item Construct a linear representation of the abstract algebra \cA\
via linear operators on some complex vector space \cV. (Because of the
structure of \cA\, the canonical commutation relations and the
anticommutation relations (if any) should, in particular, be satisfied
on \cV\.) Note that for now the $\star$-relations on \cAs\ are
to be ignored.
\item Obtain explicit operators $\hat{C}$ on \cV, representing the
quantum constraints. In general, a choice of factor ordering (and, in
the case of systems with an infinite number of degrees of freedom,
also of regularization) has to be made at this stage. {\em Physical
states} lie in the kernel \cVp\ of these operators. That is,
$\ket\psi$ is a physical state only if it is a solution to the quantum
constraint equation:
 \be\label{qp:qcon}
  \hat{C}\ket\psi=0.
 \ee
\item Extract the physical subalgebra, \cAp, of operators that leave \cVp\
invariant.  Not all operators in \cA\ are of this type. An operator
$\skhatA$ in \cA\ will leave \cVp\ invariant if and only if $\skhatA$
commutes weakly with the constraints, i.e.
 \be\label{qp:physop}
  [\skhatA, \hat{C}_I] = \sum_J \skhatf{}_I^J \hat{C}_J.
 \ee
Operators that satisfy (\ref{qp:physop}) are the quantum observables
or physical operators.  From the $\star$-relation on \cAs, we {\it
induce an involution} (denoted again by $\star$) on the physical
algebra \cAp. (See the remark in section 2.4.) Denote the
resulting $\star$-algebra of physical operators by \cAps.
\item Introduce on \cVp\ a Hermitian inner product so that the
$\star$-relations on \cAps\ are
represented as Hermitian adjoint relations on the resulting Hilbert
space.  In other words, if $\hat{F}^\star=\hat{G}$ in the abstract
algebra \cAps, then the inner product on physical states should be
chosen such that the corresponding explicit operators in the
representation satisfy $\hat{F}^\dagger=\hat{G}$, where $\dagger$ is
the Hermitian adjoint \wrt\ the physical inner product.
\end{enumerate}

Note that the steps listed above are meant to be broad guidelines
which help streamline the procedure. They are not meant to be rigid
rules. We will now discuss in some detail those features of the
program which involve some subtleties.

\subsection{Algebraic relations}
As we mentioned above, in the case when the phase space $\Gamma$ is a
genuine manifold, the completeness requirement on the space \cS\ of
elementary classical observables leads one to an {\em
over}completeness; there are algebraic relations between elements of
\cS\ . These have to be incorporated in the quantum theory in an
appropriate fashion. Let us illustrate this point by a simple example.
Consider a particle confined to a circle. Then, the configuration
space is $S^1$, the simplest non-trivial manifold, and the phase space
$T^\star(S^1)$ is topologically $S^1\times R$. Since $S^1$ does not
admit a global chart, we let \cS\ be the four-dimensional vector space
generated by, say, the set of functions $(1,\cos\theta,
\sin\theta,p_\theta)$, where $\theta$ is the standard angular parameter
on $S^1$, and $p_\theta$, its canonically conjugate momentum. Thus,
although the \ps\ is two dimensional, \cS\ contains {\it three}
nontrivial generators. Hence, in this case (in addition to the
canonical commutation relations) there is one algebraic relation to
incorporate in the quantum theory: we should require that
$(\wh{\cos\theta})^2+ (\wh{\sin\theta})^2=1$.  If we simply ignore
this relation, we would be quantizing a theory which is quite
different from the one we started out with. More generally, if the
\ps\ $\Gamma$ is $m$-dimensional and there are $n$ non-trivial
generators of \cS, one expects $n-m$ algebraic relations between the
elementary variables, which would ``cut the physical algebra down to
the right size''.

How do we incorporate the algebraic constraints in the quantum theory
in a more general setting?  Consider first the simplest case. Suppose
$F_1, ... F_n$ are elementary variables, all the Poisson brackets
between which vanish. Then, if they are subject to a relation $f(F_1,
...,F_n) = 0$ on the classical phase space, we require that
\be\label{qp:acrsim}
f(\hat{F}_1, ..., \hat{F}_n) = 0
\ee
on \cA. This condition generalizes the circle example considered
above.  Next, let us consider relations between elementary variables
whose Poisson brackets do not necessarily vanish. Now, the idea is
that if the three functions $F, G, H\equiv F\cdot G$ on $\Gamma$ are
{\it all} in the space \cs\ (and thus are to have unambiguous quantum
analogs), then we should require that in \cA:
  \be\label{qp:acr}
   \hat{F}\cdot\hat{G} +\hat{G}\cdot\hat{F} - 2\hat{H}=0 \>.
  \ee
More generally, suppose there is a set of $m$ elementary functions,
$(F_1,\dots,F_m)$, which are such that the product $ H=F_1\cdot
F_2\cdots F_m$ {\it is also an elementary variable}.  Then, the
prescription is to impose
 \be\label{qp:acrgen}
  \hat{F}_{(1}\cdot
  \hat{F}_2\cdots\hat{F}_{m)}-\hat{H} = 0,  \ee
on the algebra constructed from \cS, where as usual $(1\cdots m)$
denotes $1/m!$ times the sum of all the permutations.  Technically,
the imposition of these relations simply amounts to taking the
quotient of the free algebra by the ideal generated by the left
sides of (\ref{qp:acrsim}) and (\ref{qp:acrgen}).

Note that even though we are imposing ``anti-commutation relations'',
there is nothing fermionic about the system. Note also that the
purpose of (\ref{qp:acrsim}) and (\ref{qp:acrgen}) is {\it not} to
resolve factor ordering ambiguity since all $\hat{F}_i$ in
(\ref{qp:acrsim}) commute and since $\hat{H}$ in (\ref{qp:acrgen}) is,
by assumption, an elementary operator. Rather, (\ref{qp:acrsim}) and
(\ref{qp:acrgen}) are imposed to remove unwanted sectors in the final
quantum description.  In the circle example considered above, in
particular, the prescription rules out quantizations in which the sum
$(\wh{\cos\theta})^2+ (\wh{\sin\theta})^2$ of operators fails to equal
the identity. More generally, the algebraic identities arise due to a
``failure of global coordinatization'' of the \ps. We cannot
classically solve the algebraic identity and eliminate one of the
elementary functions {\em globally} on the \ps. The identity is then
incorporated in quantum theory through the anti-commutation relations%
\footnote{In all the examples considered in this paper as well as in
 \cite{newbook,thesis}, overcompleteness arises due to algebraic
 relations between elementary variables of the type discussed
 here. In general, however, the relations may be more complicated and
 we do not have a general prescription to handle such cases.}.

Given a classical identity $F\cdot G=H$, one might imagine
incorporating it in the quantum theory only ``up to terms of the order
$\hbar$.'' Indeed, in place of the ACR (\ref{qp:acr}), we could impose
a nonsymmetric condition of the form
 \be\label{qp:acramb}
 \hat{F}\cdot\hat{G}+\hat{G}\cdot\hat{F}+\delta\hbar\wh{\{F,G\} }
 -2\hat{H}=0.
 \ee
Since in the limit $\hbar\rightarrow 0$ the extra term on the left
side vanishes, independent of the (real) value of $\delta$ we obtain the
correct classical limit. Our prescription (\ref{qp:acr}) just sets
$\delta$ to zero.  This choice is motivated by two considerations.
First is simplicity. For definiteness, we need to give a specific
guideline to construct the $\star$-algebra of quantum operators and
setting $\delta$ to zero is the simplest choice.  Secondly, in the
standard Schr\"odinger-type quantization of systems whose
configuration space is a manifold (see, e.g., \cite{aa:cmp}, or
\cite[appendix C]{newbook}), $\delta$ {\it does turn out to be zero}.
Since this is a very large class of examples, it is reasonable to
adopt our prescription as a rule of thumb to begin with.  If in
certain systems, there are specific physical and/or mathematical
reasons {\em not} to make this choice, we can treat them as
exceptional cases and resort to the more general prescription of
(\ref{qp:acramb}).

\subsection{Selection of the inner product}
The idea behind our strategy is simply to require that real classical
observables should become self-adjoint (or, rather, symmetric) quantum
observables. In a certain sense, the idea is rather elementary. After
all, in elementary quantum mechanics, the (real) physical observables
always become Hermitian operators on the Hilbert space of quantum
states. However, the notion of using this as a general principle to
{\it select} the inner product appears to be new. As we emphasized in
section 1, the Dirac quantization program itself does not provide a
strategy to select the inner product on physical states. As a general
rule, underlying symmetries --such as the Poincar\'e invariance-- are
invoked to arrive at the inner product.  However, in many physical
systems, such as general relativity or the the finite dimensional
systems considered in this paper, there are no obvious symmetries that
can be invoked. The strategy of using the reality conditions can then
be quite powerful.

It is important to note the precise manner in which the reality
conditions are to be imposed. For simplicity of discussion let us, for
a moment, consider an unconstrained system. Then, one begins with a
vector space representation of the algebra \cA\ of operators and
invokes the reality conditions to select the inner product on the
vector space such that the $\star$-relations between abstract
operators in \cA\ become Hermitian-adjoint conditions on the concrete
operators in the {\em given} representation. A theorem due to Rendall
\cite{rendall} ensures that, if the inner product exists, then it
is unique upto an overall multiplicative constant on each irreducible
representation of \cA.  Note, however, that the procedure is tied to
the initial choice of the vector space representation. Therefore, one
can still have inequivalent quantizations. In linear field theories in
Minkowski space, for example, there are infinitely many inequivalent
representations of the CCRs in all of which the field operators are
self-adjoint. From the perspective of the reality conditions, the
inequivalence arises essentially because of the freedom in the initial
vector space representation.

Note that there is no a priori guarantee that the vector space would
admit an inner product satisfying the required quantum reality
conditions.  If it does not, one must change the vector space
representation judiciously and start all over again.  Thus, the
success of the procedure does depend on the initial choice.  A simple
example in which no inner-product implementing the reality conditions
exists in an ``obvious'' vector space representation, but can be found
uniquely in a slightly modified one, is discussed in \cite[p.\
155]{newbook}).

Perhaps the most interesting examples of the success of this strategy
are provided by toy models that arise from general relativity:
mini-superspaces (see, e.g., \cite{atu:II,gamm,gmt:b2}) and
2+1-dimensional gravity (see, e.g., \cite[\S 17]{newbook}). Let us
focus on 2+1 gravity because it shares all the conceptual difficulties
with the 3+1-dimensional theory. In this case, it was generally
believed that it would be impossible to find the physical inner
product without ``deparametrizing'' the theory first (i.e., separating
an internal time variable from the true, dynamical degrees of
freedom). In the general case when the Cauchy slices are of a genus
greater than one, the problem of deparametrization is still open.
However, the strategy of using the reality conditions does work and
selects the inner product unambiguously (in the connection
representation, see e.g.\ \cite[\S 17]{newbook}). Thus, the strategy
enables one to separate the problem of time from the problems of
finding the complete mathematical structure, including the Hilbert
space structure on the space of physical states. In full,
3+1-dimensional quantum gravity, it may well be that time can be
introduced only as an approximate notion.  It is therefore important
to have alternate strategies available to construct the complete
mathematical framework.

In this paper, we will consider only finite dimensional examples and
will need to introduce the inner product just on physical states,
i.e., solutions to constraints. Indeed, even if one
introduced an inner product on the space \cV, solutions to
constraints will typically not be normalizable with respect to this inner
product. In certain cases --especially for systems with an infinite
number of degrees of freedom-- it is nonetheless useful to
have an inner product available on \cV\ for technical reasons: this
structure can restrict the factor ordering choices, enable one to
regulate products of operator-valued distributions that may feature in
the expressions of constraints and naturally suggest the appropriate
function spaces in which the solutions to constraints should lie
\cite{haj}. In such cases, then, one can use the reality conditions
on \cV\ to introduce a fiducial, kinematic inner product.

\subsection{Physical $\star$-algebra}
Let us explore the space of physical operators. An operator $\hat{A}$
in \cA\ will leave the space of solutions to the quantum constraint
equation (\ref{qp:qcon}) invariant if and only if $[\hat{A},
\hat{C}_I] = \sum_J \skhatf{}_I^J \hat{C}_J$. The collection of all such
operators --which includes the constraints themselves-- forms a
sub-algebra of \cA. Denote it by \cAp$'$. Since constraints annihilate
all physical states, the algebra \cAp\ of physical observables can be
obtained ``by setting the constraint operators to zero'' in \cAp$'$.
More precisely, one constructs the ideal ${\cal I}_C$ of \cAp$'$
generated by the constraints and takes the quotient of \cAp$'$ by this
ideal: \cAp:=\cAp$'/ {\cal I}_C$.  Since there may be some ambiguity
in the correspondence between the classical and quantum observables,
it is at this stage that some factor ordering choices for the
physical operators may have to be made, before one quotients by
${\cal I}_C$.

Now, if $\hat{A}$ in \cA\ belongs to \cAp$'$, its $\star$-adjoint in
\cA, $\hat{A}^\star$, may not belong to \cAp$'$. Hence, in general the
$\star$-relation on \cA\ does {\em not} induce an involution on \cAp.
In this case, no prescription is available to select the physical
inner product in the simple program outlined in section 2.1. If, on
the other hand, $\hat{A}^\star\in\hbox{\cAp}\>\forall\>
\hat{A}\in\hbox{\cAp},$ then we do obtain an involution on \cAp\
(denoted again by $\star$) and hence a physical $\star$-algebra \cAps,
which can now be used to select the inner product. For what kind of
physical systems is the above condition guaranteed to be satisfied?
Consider, for example, the case when the constraint operators satisfy:
$\hat{C}_I{}^\star=\hat{C}_I,\>\forall I$ (i.e.\ the classical
constraint functions are all real). Now, if a physical operator
$\hat{A}$ commutes strongly with the constraints, i.e.\ if
$[\hat{A},\hat{C}_I]=0$, then $\hat{A}{}^\star$ is also a physical
operator, even though $\hat{A}^\star\neq\hat{A}$. We will see that
this situation occurs in a number of model systems.  In these cases,
the quantization program can be completed successfully. Note however,
that it is {\em not essential} that the operators commute strongly
with the constraints for the $\star$-relations to be well-defined on
the physical algebra. For example, if a set of generators of \cAp\ are
their own $\star$-adjoints (i.e.\ if they correspond to {\it real}
functions on the \ps), the $\star$-relations on \cA\ induce an
unambiguous $\star$-relation on \cAp.

Finally, note that to capture the full physics of the model under
investigation, the algebra \cAp\ has to be complete in an appropriate
sense: the set of classical functions corresponding to the generators
of \cAp\ should be {\em complete} on the reduced \ps. In terms of the
constraint surface itself, this amounts to the requirement that the
set of Hamiltonian vector fields of the classical observables
corresponding to the generators of \cAp$'$ span the tangent
space to the constraint surface. In practice, it is often the case
that, if the completeness condition is not met globally, the theory
has superselection rules; the ``obvious'' vector space representations
are not irreducible.

\subsection{Remarks}
We will conclude with a few remarks.
\medskip

\noindent{\sl Completeness of \cVp:}

Is there a criterion which will determine whether the space of
solutions to the quantum constraint equation is physically ``large
enough''? Since --after one has found an inner product-- \cVp\ is
complete as a Hilbert space (and any vector space with a norm can be
completed) this notion of completeness cannot help resolve the above
question. However, we do have a physical notion of completeness of
\cAp, and thus we can require that \cVp\ be large enough to carry a
{\em faithful representation} of the algebra of observables.
\medskip

\noindent{\sl Solutions and physical states:}

Physical states are {\it normalizable} solutions to the quantum
constraint equation. However, not all solutions to the quantum
constraints are normalizable with respect to the physical inner
product. Thus, there is a distinction between physical states and
solutions. Of course, physical states form a (generically proper)
subspace of the kernel of the constraint operator.  While one must
keep this distinction in the back of one's mind, we will no longer
emphasize it.
\medskip

\noindent{\sl Inputs to the program:}

As we have emphasized, the steps outlined in section 2.1 do not
constitute a crank that can be turned to convert a constrained
classical theory to a quantum theory. Rather, the steps constitute
broad guidelines that clarify and streamline the choices that are
available in the transition. These choices arise through three main
inputs: {\sl i)} the choice of the space \cS\ of elementary classical
variables; {\sl ii)} the choice of the vector space representation of
the algebra \cA; and, {\sl iii)} the choice of factor ordering in the
expression of constraints. While these choices {\em are} restricted by
a number of consistency conditions, there is nonetheless considerable
freedom left at the end. In making these choices, therefore, one must
use physical input.  Generally, the final quantum theory one obtains
{\em does} depend on the choices in quite a sensitive manner.  This is
not surprising: since the goal of quantization is to obtain the more
complete quantum description starting from the ``coarse-grained''
classical theory, it necessarily requires a certain amount of new
information which is hard to specify universally at the outset.

A particularly natural avenue to making these choices is provided by
the group theoretic approach developed by Isham \cite{isham:lh}. If
the appropriate ``canonical group'' is identified, it automatically
provides the elementary variables, the operator expressions of the
constraints, {\it and} the inner product on physical states. (For
interesting examples, see, e.g., \cite{isham:lh, cr:prd}.) Thus,
this approach succinctly reduces the problem of finding the inputs
needed in the algebraic program to that of finding an appropriate
group action on the phase space. Another avenue is provided by
geometric quantization: If a polarization which is appropriately
compatible with the constraints can be found, one is led to physical
states and a complete set of physical observables (see, e.g.,
\cite{aa:ms}). One must still find, however, the inner product on the
physical states via, e.g., the reality conditions.

\mysection{Harmonic oscillator in the $(q,z)$ variables}

We now present the simplest application of the quantization program.
The aim is only to illustrate the role of the reality conditions.
Therefore, in this section, we will ignore constraints.

As we discussed in the introduction, several features of our extension
of the Dirac quantization scheme have been motivated by peculiarities
of general relativity. One of these is the fact that the Hamiltonian
formulation simplifies if one uses a ``hybrid'' pair of canonical
variables (see e.g. \cite{AA:87,newbook}), where one variable is real
and the other is complex. In quantum theory it is then simplest to
work in a representation in which the states are holomorphic
functionals of the complex variable, treat the real variable as
momentum and represent it by a functional derivative
\cite{AA:87,newbook}.  A question that arises immediately is whether
the procedure is consistent: can the momentum conjugate to a complex
variable be itself Hermitian?

To address this issue in a simple context, in this section we will
consider the harmonic oscillator in terms of hybrid variables.  More
precisely, we will quantize the oscillator using hybrid elementary
variables and find that the implementation of the Hermiticity
conditions leads to a quantum theory which is unitarily equivalent to
the usual one.  Thus there is no {\em a priori} obstruction to using
such variables.

Consider the \ps\ of a harmonic oscillator of unit mass and spring
constant.  $\Gamma$ is coordinatized by the real canonically conjugate
functions $(q,p)$. In analogy with \gr\ in the connection variables, let us
introduce the complex variable $z=q-ip$. Note that the \ps\ is real;
$(q,z)$ do not constitute a chart on $\Gamma$.  However, we choose as
\cS\ the vector space spanned by the (complex) functions $(1,q,z)$.
This set is complete, and closed under Poisson brackets. \cS\ is
also closed under complex conjugation, since $\bar{q}=q$ is real and
$\bar{z}=2q-z\in\cS$. These hybrid reality conditions are analogous to
those for the new variables in linearized \gr. The canonical
commutation relations between the corresponding operators are
 \be \label{qz:ccr}
  [\hq,\hz]=\hat{1},
 \ee
and the $\star$-relations are:
 \be \label{qz:star} \hq^\star=\hq,\quad\hz^\star=2\hq-\hz.
 \ee
(These are completely analogous to those one encounters if one
linearizes general relativity in the hybrid canonical variables
\cite[section 11.5]{aa:jl,newbook}.)  We next have to find a
representation of this algebra. The hybrid variables suggest a new
approach. We can represent the above operators on the space of {\it
holomorphic} functions of one complex variable.  Let \cV\ be the space
of holomorphic functions $\psi=\psi(z)$, and represent the operators
by:
 \be\label{qz:rep}
  \hz\circ\psi(z)=z\cdot\psi(z)\quad\hbox{and}\quad
  \hq\circ\psi(z)= \frac{d}{dz}\psi(z).
 \ee
For linearized gravity, and indeed for \gr\ itself, the analogous
holomorphic (or self-dual) connection representation is particularly
convenient since it greatly simplifies the form of the constraints.

A natural ansatz for the inner product on these holomorphic states is
 \be\label{qz:ip}
  \IP\psi\chi=\lint{dz\wedge d\bar{z} \over2\pi
   i}\>e^{\mu(z,\bar{z})}\>\bar\psi\,\chi,
 \ee
where $\mu=\bar\mu$ is a function to be determined. We now have to
impose the Hermiticity conditions (\ref{qz:star}) on these operators.
This is the crucial step: it is quite counterintuitive to have a real
momentum (represented by a holomorphic derivation) ``canonically
conjugate'' to a complex variable, and hence the consistency of the
formalism is not immediately obvious.

However, the calculation is straightforward.  The Hermiticity
condition on $\hq$ yields a differential equation for $\mu$ which
constrains $\mu$ to be of the form $\mu=\mu(z+\bar{z})$. Next, using
the holomorphicity of the wavefunctions, the Hermiticity condition on
$\hz$ yields another differential equation for $\mu$ which is solved
by:
 \be\label{qz:meas}
  \mu(z,\bar{z})=-{(z+\bar{z})^2\over4}.
 \ee
Hence, the Hermiticity conditions do determine the measure uniquely.
Note that even though the measure does not ``fall-off'' as expected
(for example in the $Im(z)$ direction), there exist normalizable
states: these are of the form $\psi(z)=e^{z^2\over4}f(z)$, where
$f(z)$ are polynomials in $z$. The inner product
(\ref{qz:ip},\ref{qz:meas}) now yields:
 \be\label{qz:bargip}
   \IP\psi\psi=\lint{dz\wedge d\bar{z} \over2\pi
   i}\>e^{-\frac{z \bar{z}}2}\>|f(z)|^2.
 \ee

How is this quantization related to the Bargmann quantum theory? Apart
from a factor of $2$ in the exponent which arises because the $z$ we
have defined is $\sqrt2$ times the usual Bargmann variable, the inner
product (\ref{qz:bargip}) corresponds to the standard Gaussian measure
on the Bargmann states, establishing a unitary map between the two
Hilbert spaces.  Next, we can use this unitary map to compare the
actions of the operators $(\hq,\hz)$ in the two representations. The
result is that the two quantum theories are unitarily equivalent%
\footnote{Alternately, one could represent $\hq$ by $\hq\circ\psi(z)
 = \frac1{\sqrt2}({\displaystyle{\sqrt2}}\frac{d}{dz} +
 \frac{z}{\sqrt2}) \psi(z)$.
 Then, the commutation
 relations are still satisfied, but the measure $\mu(z, \bar{z})$ is
 the usual Bargmann
 measure $e^{-\frac{z\bar{z}}{2}}$. The unitary equivalence is then
 manifest. (Recall that $z/\sqrt2$ is the usual Bargmann variable.)}.

Thus, we have completed the quantization of the 1-dimensional
oscillator in the hybrid $(q,z)$ variables. The Hermiticity conditions
on the elementary operators can be implemented, and they fix the inner
product. This indicates that there is {\it a priori} no obstruction to
the use of a hybrid set of variables of the type used in \gr.
\medskip

\noindent{\sl Remark:}

As we noted after Eq.(\ref{qz:star}), the variables $(z,q)$ we used
above and the resulting reality conditions are completely analogous to
those encountered in the linearized version of general relativity. In
the exact theory, however, one of the reality conditions is {\em
non-polynomial}.  To mimic this situation, Kucha\v r
\cite{kk:ex} has proposed the following model. Let the phase
space be 2-dimensional, labelled by the real canonically conjugate
variables $(Q,P)$, with $Q>0$, and let the
Hamiltonian be $h = QP^2 +1/Q$.  Then, if we set $Z= 1/Q -iP$, and
treat $Q,Z$ as the hybrid canonical variables, the Hamiltonian
simplifies to $h= 2Z -QZ^2$; as in general relativity, it becomes a
low order polynomial in the new variables.  However, now one of the
reality conditions is non-polynomial: $\bar{Z} = -Z +2/Q$. It was
thought that its complicated nature would be an unsurmountable
obstacle in carrying out the quantization program and concern was
expressed that the situation may be similar in full general relativity
\cite{kk:ex}.

It turns out however, that, by appropriately choosing a vector space
representation of the algebra $\cA$ and an ansatz for the inner
product, the quantization program can be completed in the
Kucha\v r model too. The final description is as follows. We can choose
$(1, Q, QZ)$ as the elementary variables.  The $\star$-relations
are: $Q^\star =Q$ and $(\widehat{QZ})^\star = 2 - \widehat{(QZ)}$. Let us
choose for the vector space \cV\ the space of holomorphic functions
$\Psi(z)$ of one complex variable $z =x+iy$, and set:
\bea\label{kk:hops}
\widehat{Q}\circ \Psi (z) &=& \left( \frac{d}{dz}+\frac{z}2\right)
\Psi (z) \mbox{ and}\\
\widehat{QZ}\circ \Psi (z) &=& \left( z\frac{d}{dz}+\frac{z^2}2+
\frac32\right)\Psi (z),
\eea
It is easy to check that the CCR are satisfied; we have a vector space
representation of \cA.  The inner product which implements the
$\star$-relations is:
\be\label{kk:ip}
\left\langle \Psi |\Phi \right\rangle =\int dy\mbox{ }e^{-\frac12 y^2}
\left.\left( \overline{\Psi \left( z\right) }\Phi \left( z\right) \right)
\right|_{x=0},
\ee
where we have used the fact that holomorphic functions are completely
determined by thier restriction to the $x=0$ line. The Hamiltonian
which generates the quantum dynamics is given by:
\be\label{kk:ham}
\widehat{h}\circ \Psi (z) = -\left( z^2\frac{d}{dz}+
\frac{z^3}2+z\right) \Psi (z).
\ee
It is straightforward to check that all three operators defined above
are Hermitian and their commutators are $i$ times the Poisson brackets
of their classical analogs. Thus, the non-polynomiality of the reality
conditions is not necessarily an obstacle to the completion of the
program.

\mysection{Coupled Oscillators}

In the previous section we quantized an unconstrained system and used
the (nontrivial) Hermiticity conditions on the elementary variables to
determine the inner product on quantum states. In this section we will
consider two constrained systems, and illustrate how one can use the
Hermiticity conditions on {\em physical observables} to obtain an
inner product on physical states. Each of these systems will be built
out of two harmonic oscillators with the same frequency, which will be
set to 1 for simplicity.

In the more interesting of the two models, the oscillators are coupled
to each other via a first class constraint on the {\em difference}
between their energies. This model is of interest especially because
it mimics certain features of \gr\ in the geometrodynamical variables:
{\it i}) the constraint is quadratic in momenta; {\it ii}) the kinetic
piece of the constraint is of indefinite signature; and, {\it iii})
the potential term is also of indefinite sign. Due to these
similarities, some of the qualitative results are of interest to \qg,
particularly quantum cosmology. In fact, this specific model
corresponds precisely to the Friedman-Robertson-Walker universe (with
$S^3$ spatial topology), conformally coupled to a massless scalar field
(see section 4.3).

However, we will begin with a simpler but related model in which the
two oscillators are coupled to each other via a first class constraint
on the {\em sum} of their energies.  This model has been used in the
past as a testing ground for various ideas (e.g.\ the issue of time
\cite{narain,cr:time}). Since it is simpler than the ``energy
difference'' model, it will allow us to implement some of the new
features of the quantization program in a familiar setting.  We will
then use the same approach for the energy difference model.

In the first two subsections, we will construct the Dirac quantum
theories for the two models. (The reduced space quantum theories have
been constructed in \cite{thesis}.) In the last subsection, we discuss
several features of the energy difference model, including its
relation to the minisuperspace mentioned above.

\subsection{Constrained total energy}
The 4-dimensional \ps\ of the system is described by position and momentum
coordinates $(x_I,p_I,\>I=1,2)$. The first class constraint we wish to
impose is
 \be \label{sum:con}
  C\equiv\half(p_1^2+x_1^2+p_2^2+x_2^2)-E\approx0,
 \ee
where $E\ge0$ in order for the classical system to be well defined.
Choose as the set \cS\ of elementary classical variables the standard
``creation'' and ``annihilation'' functions on $\Gamma$
 \be\label{sho:var}
   z_I={\textstyle{1\over\sqrt2}}(x_I-ip_I)\quad\hbox{and}\quad
        \bar{z}_I= {\textstyle {1\over\sqrt2}}(x_I+ip_I),
 \ee
as well as the constant function. Since these elementary variables are
algebraically independent, there are no ACRs in the quantum algebra \cA.
In these variables, the constraint function is
 \be\label{sum:zcon}
    C=(z_1\bar{z}_1+z_2\bar{z}_2)-E .
 \ee

The quantum $\star$-algebra \cAs\ is straight forward to construct.
To make the notation transparent we will denote the elementary quantum
operators $\hat{z}_I$ by $\hc_I$ and $\hat{\bar{z}}_I$ by $\ha_I$.
\cAs\ is then generated by the set of elementary quantum operators
($1,\ha_1,\hc_1,\ha_2,\hc_2$) which satisfy the canonical commutation
relations:
 \be \label{sho:ccr}
  [\ha_I,\ha_J]=0=[\hc_I,\hc_J]\quad\hbox{and}\quad [\ha_I,\hc_J]=
         \delta_{IJ},\quad I,J=1,2;
 \ee
and are subject to the $\star$-relation
 \be\label{sho:elemstar} \ha_I^\star=\hc_I,\qquad\forall I.
 \ee
 In terms of these operators, the quantum constraint we wish to impose
is
 \be\label{sum:qcon}
  \hat{C}\ket{\psi}_{phy}:=\left(\hc_1\ha_1
         +\hc_2\ha_2+1-E\right)\ket{\psi}_{phy}=0,
 \ee
where we have symmetrized the operator product to resolve the ordering
ambiguity in the constraint. (See the remark at the end of this section.)

The next step in the quantization program is to represent the algebra
\cA\ by means of concrete operators on a vector space \cV.  (Recall
that the $\star$-relations are ignored at this stage.)  Let us choose
the vector space representation of \cA\ as follows. Since any complete
set of commuting operators consists of only two of the elementary
operators, let us choose for \cV\ the complex vector space spanned by
states of the form $\ket{j,m}$, where (to begin with) $j$ and $m$ are
any {\it complex} numbers, and let us represent the elementary quantum
operators as follows:
 \bea \label{sho:rep}
  \ha_1\ket{j,m} &=& \alpha_1(j+m)\ket{j-\half,m-\half}, \cr
                \hc_1\ket{j,m} &=& \gamma_1(j+m+1)\ket{j+\half,m+\half}, \cr
                \ha_2\ket{j,m} &=& \alpha_2(j-m)\ket{j-\half,m+\half} \cr
   \hbox{and}\quad
  \hc_2\ket{j,m}&=&\gamma_2(j-m+1)\ket{j+\half,m-\half};
 \eea
where the coefficients, $\alpha_I(k)$ and $\gamma_I(k)$, functions only
of their argument $k$, are chosen to satisfy
 \be  \label{sho:cond}
  \alpha_1(k)\gamma_1(k)=k \quad\hbox{and}\quad\alpha_2(k)\gamma_2(k)= k.
 \ee
It is straightforward to check that the commutation relations
(\ref{sho:ccr}) are satisfied by any representation (\ref{sho:rep}) in
which (\ref{sho:cond}) is satisfied. Thus, so far we have a (rather
large) family of vector space representations and none of them is
preferred. Eq.\ (\ref{sho:rep}) implies that the number operators
$\hN_I=\hc_I\ha_I$ are represented simply by
$\hN_1\ket{j,m}=(j+m)\ket{j,m}$ and $\hN_2\ket{j,m}=(j-m)\ket{j,m}$.
Each $\ket{j,m}$ is an eigenket of the total number operator $\hN
=\hc_1\ha_1 +\hc_2\ha_2$ with eigenvalue $2j$, as well as the
difference between the number operators $\hN_1-\hN_2$ with eigenvalue
$2 m$. (Thus, had we represented states as Bargmann type holomorphic
wave functions, we would have $\psi(z_1, z_2):=\IP{z_1,z_2}{j,m}\equiv
z_1^{j+m}z_2^{j-m}$.) The notation $\ket{j,m}$ to represent the kets
may seem strange at first.  However, these angular momentum like
states arise naturally because, as we will see below, the Poisson
bracket algebra of {\em physical} observables is the Lie algebra of
$SO(3)$.  The $\ket{j,m}$ representation will therefore be more
convenient than the number representation.

Since $\hC$ is diagonal in this representation, with eigenvalues
$2j+1-E$, the quantum constraint is easy to solve.  A basis for the
space \cVp\ of solutions to the quantum constraint equation is given
simply by the kets $\{\ket{j=L,m}\}$, where $L$ is now a fixed number,
$L=\frac{E-1}{2}$.  Note that this is a ``small'' subspace of \cV; $j$
was an arbitrary complex number to begin with, but is now a {\it fixed
real number}. However, $m$ is still allowed to be an arbitrary complex
number.  Physical operators are the elements of ${\cal A}$ that map
\cVp\ to itself; their action should preserve the total energy of the
two oscillators. Clearly, operators that simultaneously raise the
energy of one and lower the energy of the other oscillator by a unit,
and an operator that measures the energy difference, are physical
operators.  Hence, let us consider the algebra generated by the set of
operators $\{\hL_z, \hL_\pm\}$, where $\hL_z:=\half(\hN_1-\hN_2)$,
$\hL_+:=\hc_1\ha_2$ and $\hL_-:=\hc_2\ha_1$ are physical operators.
The commutator algebra is given by
 \be\label{sum:occr}
  [\hL_+,\hL_-]=2\hL_z \quad\hbox{and}\quad [\hL_z,\hL_\pm]=\pm\hL_\pm ;
 \ee
it is isomorphic to the Lie algebra of $SO(3)$. The Hilbert space of
physical states will thus provide a unitary representation of $SO(3)$.

These representations are of course well known.
However, we will arrive at them systematically following the various
steps in the quantization program. This procedure will also help
prepare the reader for our next example where the representation
theory of the observable algebra is not so well known.

Using (\ref{sho:rep}), we can evaluate the action of the physical
operators on the physical states. Doing so, we get
 \bea  \label{sum:orep}
  \hL_z\ket{L,m} &=& m\ket{L,m}, \cr
             \hL_+\ket{L,m} &=& \lambda_+(m+1)\ket{L,m+1} \cr \hbox{and}
        \quad\hL_-\ket{L,m} &=& \lambda_-(m)\ket{L,m-1},
 \eea
where $\lambda_\pm$ ---functions of their arguments only---
are just products of the coefficients $\alpha_I, \gamma_I$ in (\ref{sho:rep}).
Since $\alpha_I,\gamma_I$ are
(non-unique) solutions of
(\ref{sho:cond}), the coefficients $\lambda_\pm$ satisfy
 \be   \label{sum:ocond}
  \lambda_+(m) \lambda_-(m)= (L+\half)^2-(m-\half)^2. \ee
Since only the $\lambda_\pm$ are relevant to the observable algebra,
we can view (\ref{sum:ocond}) directly as a {\it condition} on
$\lambda_\pm$.  With this condition, the canonical commutation
relations of the observable algebra are also identically satisfied.
Recall that we had considerable freedom in our choice of the
representation (\ref{sho:rep}) of the operator algebra \cA\ since the
coefficients $\alpha_I$ and $\gamma_I$ were arbitrary to a large
extent. This freedom descends to the {\it physical} operator algebra:
due to the existence of a multitude of solutions to (\ref{sum:ocond}),
the representation (\ref{sum:orep}) of the physical algebra is also
not unique.

Recall that the program requires us to construct a {\em complete}
algebra of observables. Let us pause to analyze this issue.
The classical analogs of $\{\hL_z,\hL_\pm\}$ are the functions
 \be  \label{sum:obs}
  L_z:=\half(z_1\bar{z}_1-z_2\bar{z}_2),\quad L_+:=z_1\bar{z}_2,\quad
       L_-:=\bar{z}_1z_2
 \ee
on \ps%
\footnote{We have denoted these observables by $(L_z,L_\pm)$ rather
than $(J_z,J_\pm)$ only to distinguish them from a related but
different set of functions ---that serve as observables in the next
example--- which will be denoted by $(J_z,J_\pm)$. No relation with
orbital (as opposed to total) angular momentum is intended.}.
One can easily check that the set $(L_+,L_-)$ is by itself
{\em complete}, the set suffices to coordinatize the reduced \ps. It
is in order to ensure that the algebra of observables is {\em closed
under Poisson brackets} that one has to include $L_z$ in the set of
generators of \cAp, and thus make the set {\em over}complete. There is
an algebraic relation satisfied by this overcomplete set which, it
turns out, fixes a value of the Casimir invariant of \cAp.  Using the
definitions of $\hL_\pm$ and the commutation relations
(\ref{sho:ccr}), one finds that the algebraic relation is
$\hL^2:=\hL_z^2+\half[\hL_+,\hL_-]_+=L(L+1)$. (Recall that the value
$L=\frac{E-1}2$ is determined by the classical constraint.)
Equivalently, the relation can be expressed as
 \be\label{sum:oacr}
  \hL_+\hL_-=(L+\half)^2-(\hL_z-\half)^2,
 \ee
in which form it is manifest that this condition on the operators is
automatically satisfied because of (\ref{sum:ocond}). Thus the
completeness requirement has been incorporated in our representation.

The next step in the program is to obtain an inner product on the
space of physical states by requiring that the $\star$-relations on
the physical operators become Hermitian adjointness relations on the
resulting Hilbert space. From the expressions for the physical
operators in terms of the elementary quantum operators $\ha_I,\hc_I$
and the $\star$-relation (\ref{sho:elemstar}), one obtains the
$\star$-relation induced on
\cAp:
 \be  \label{sum:o*}
  \hL_+^\star=\hL_- \quad\hbox{and}\quad \hL_z^\star=\hL_z.
 \ee
First, consider only the operator $\hL_z$. Since it is to be Hermitian
on the physical Hilbert space, its eigenvalues must be real, and its
eigenkets with distinct eigenvalues must be orthogonal to each other.
Hence $m$ is real. (Note that $L$ is already real on all solutions to
the constraint.) Next, recall that the Hermiticity conditions are to
be implemented separately on each {\it irreducible} representation of
the algebra. The representation (\ref{sum:orep}), however, is
reducible: the physical operators either leave the value of $m$
unchanged, or change it by an {\it integer}. Thus, the fractional part
of $m$ ---denoted by $\epsilon={\rm frac}(m)$--- is invariant under
the action of $\hL_z,\hL_\pm$. Now, consider ${\cal
V}_{phy}^\epsilon$, the vector space of states with the same fixed
value of $\epsilon$. {\em Each subspace ${\cal V}_{phy}^\epsilon$
carries an irreducible representation of the $SO(3)$ Lie algebra}
(\ref{sum:occr}). Let $m=n+\epsilon$, $n=\cdots -2,-1, 0,1,2\cdots$.
Each ${\cal V}_{phy}^\epsilon$ has a countable basis, the elements of
which are labelled by $n$ ---the integer part of $m$--- and it is on
these {\it irreducible} representations that we now wish to implement
the remaining Hermiticity conditions. Prior to this implementation, we
have a 1-parameter family of ambiguities in the quantization of the
system, labelled by the parameter $\epsilon\in [0,1)$.

For definiteness, consider a representation with a fixed
value of $\epsilon$. The Hermiticity of $\hL_z$ implies that on ${\cal
V}_{phy}^\epsilon$ there exists an inner product in which the above
basis is orthogonal; without any loss of generality, we can choose it
to be orthonormal. Thus the inner product can be chosen to be:
 \be   \label{sum:ip}
  \IP{L,m'=n'+\epsilon}{L,m=n+\epsilon}=\delta_{n',n},
 \ee
where both states on the left have the same fractional part of $m$.
Note that it is only because we implement the Hermiticity conditions
on an {\em irreducible} sector ---with a countable basis--- that we
can postulate a Kronecker-$\delta$ normalization on ${\cal
V}_{phy}^\epsilon$.  Had we tried to introduce an inner product on the
entire \cVp, we would have been led to a Dirac-$\delta$ normalization.
Finally, as in the familiar quantization of the $SO(3)$ algebra, it is
straightforward to show that the Hermiticity conditions (\ref{sum:o*})
fix the coefficients $\lambda_\pm(m)$.  One finds that there exist
non-trivial representations only when $\epsilon=\rmfrac(L)$, and then
only when $L$ itself is half integer or integer (and thus {\em for
integer $E$ only}). The representation of the generators of
\cAp\ is
 \bea  \label{sum:orepf}
  \hL_z\ket{L,m} &=& m\ket{L,m}, \cr
             \hL_+\ket{L,m} &=& \sqrt{(L-m)(L+m+1)}\ket{L,m+1} \cr \hbox{and}
        \quad\hL_-\ket{L,m} &=& \sqrt{(L+m)(L-m+1)}\ket{L,m-1},
 \eea
where as before $L={(E-1)\over2}$. It is easy to check that in the
inner product (\ref{sum:ip}), the Hermiticity conditions
$\hL_z^\dagger=\hL_z$ and $\hL_-^\dagger=\hL_+$ are satisfied,
implementing the $\star$-relations.  This representation
(\ref{sum:orepf},\ref{sum:ip}) of the observable algebra is unique
up to unitary equivalence, since the Hermiticity conditions can only be
implemented on the subspace labelled by $\epsilon=\rmfrac(L)$. Not
surprisingly (since the reduced \ps\ is compact), the representation
is finite-dimensional, of dimension $2L+1$. A basis is provided by
states with $m=-L,-L+1,\cdots L-1,L$. In the final analysis, the
representation we have obtained is not surprising, given that \cAp\ is
just the Lie algebra of $SO(3)$. However, we took this long route to
to show that the quantization program, when implemented step by step,
does lead one to the expected result.

Note that the coefficients $\alpha_I, \gamma_I$ in (\ref{sho:rep}) are
left undetermined, and we have {\it not} obtained an inner product on
the original representation space. However, (\ref{sum:ip}) provides us
with an inner product on \cVp. Let us summarize the steps by which the
reality conditions have led us to the final result. Prior to imposing
the reality conditions, there was considerable freedom in the choice
of representation of the physical algebra: there exist representations
of the observable algebra on solutions to the constraint labelled by
{\em complex}-valued $m$, the coefficients $\lambda_\pm$ are not
unique, and for each choice of $\lambda_\pm$, there is a one parameter
family of irreducible representations, labelled by
$\epsilon$. However, not all these representations are compatible
with the reality conditions. The Hermiticity condition on one of the
physical operators, $\hL_z^\dagger=\hL_z$, allows only representations
on states labelled by real $m$, with the inner product
(\ref{sum:ip}). The further reality condition, $\hL^\dagger_+=\hL_-$,
then implies that there exist representations only for integer values of
$E$, and then the representation is unique.

The representations we have constructed above are only for integer
$E$. Is it possible to construct representations for arbitrary
(positive) $E$?  For the factor-ordering of the constraint that we
have chosen, there is {\em no quantum theory} for non-integral $E$.
{}From the point of view of geometric quantization, this is not
surprising since the reduced \ps\ is a compact manifold, $S^2$, and it
admits a K\"ahler structure only for integer values of the radius
\cite{njmw,ms:thesis}.  From ordinary quantum mechanics, we are familiar
with the fact that the commutation relations (\ref{sum:occr}) have
only the integer and half-integer spin representations.  Note however
that there is a factor-ordering ambiguity in the choice of constraint
operator. If we allow $z_I\bar{z}_I$ to be represented by an
undetermined convex linear combination of $\hc_I\ha_I$ and
$\ha_I\hc_I$, then the resulting quantum constraint equation is
 \be\label{sum:amb}
  \hat{C}\ket\psi_{phy}\equiv(\hc_1\ha_1+\hc_2\ha_2+1+\kappa-E)\ket\psi_{phy}
      =0,\quad
 \ee
where $\kappa\in[-1,1]$ represents the factor-ordering ambiguity. Now,
even though $L=\frac{E-\kappa-1}2$ is still integer or half-integer,
$E$ is no longer forced to be integer. Thus, there are (two) choices
of ordering of the quantum constraint operator which allow us to find
a non-trivial physical quantum theory, even when $E$ is not an
integer.

\subsection{Constrained energy difference}

The unconstrained \ps\ $\Gamma$ for this model is again ${\rm I}\!{\rm
R}^4$, coordinatized by real variables $(x_I,p_I)$ or complex
variables $(z_I,\bar{z}_I)$, $I=1,2$.  The first class constraint we
now wish to impose is however different.  We will require that the
difference between energies be a constant, $\delta$:
 \be\label{dif:zcon}
  C:=\half\left(z_1\bar{z}_1-z_2\bar{z}_2\right)-\delta\approx0.
 \ee
We will use the elementary operators
(\ref{sho:ccr}-\ref{sho:elemstar}) and representation (\ref{sho:rep})
defined in the previous section.  In terms of these operators, the
quantum constraint we wish to impose is
 \be  \label{dif:qcon}
  \hat{C}\ket{\psi}_{phy}:=\left[\half(\hc_1\ha_1
  -\hc_2\ha_2)-\delta\right]\ket{\psi}_{phy}=0,
 \ee
where,
since the constraint is the {\it difference} between the energies of the two
oscillators, as long as we use the same ordering for each term
$z_I\bar{z}_I$, there is no ambiguity in the constraint operator.

Since $\hC$ is diagonal in this representation,
with eigenvalues $m-\delta$, the quantum constraint is easy to solve.
A basis for the {\it physical} subspace
\cVp\ is given simply by the kets $\{\ket{j,m=\delta}\}$, where $j$ is
an arbitrary complex number. Note that the situation here is reversed
from the previous example: there, the constraint forced $j=L$ to be
real and left $m$ arbitrary.

Next let us consider physical operators: these are elements of ${\cal
A}$ that map \cVp\ to itself and should thus maintain the difference
in energies of the two oscillators.  Clearly, operators that raise and
lower the energy of each oscillator by a unit, and an operator that
measures the total energy, are physical operators. Hence, consider the
algebra generated by the set $\{\hJ_+,\hJ_{-},\hJ_z\}$; where
$\hJ_+:=\hc_1\hc_2$ raises the energy of each oscillator by a unit,
$\hJ_-:= \ha_1\ha_2$ lowers the energy of each oscillator by a unit,
and $\hJ_z:=\half(\hN+1)$ is half the total energy.  The commutation
relations between these operators are
 \be  \label{dif:occr}
  [\hJ_+,\hJ_-] =-2\hJ_z \quad\hbox{and}\quad [\hJ_z,\hJ_\pm]=\pm\hJ_\pm ,
 \ee
from which it is clear that they generate the Lie algebra of
$SO(2,1)$. (Note the difference from (\ref{sum:occr}) in the sign of
the first commutator.)  $\hJ_+$ and $\hJ_-$ are the (angular
momentum) raising and lowering operators, respectively.  Note also
that it is in order to ensure the closure of the commutation relations in
(\ref{dif:occr}) that we have chosen the definition $\hJ_z:=
\half(\hN+1)$, as opposed to $\hJ_z=\half\hN$.

Using (\ref{sho:rep},\ref{sho:cond}), we can evaluate the action of
the physical operators on the physical states to obtain
 \bea \label{dif:orep}
  \hJ_z\ket{j,\delta} &=& (j+\half)\ket{j,\delta}, \cr
              \hJ_+\ket{j,\delta} &=& \kappa_+(j+1)\ket{j+1,\delta} \cr
    \hbox{and}\quad
  \hJ_-\ket{j,\delta}&=&\kappa_-(j)\ket{j-1,\delta},
 \eea
where $\kappa_\pm$ ---functions of their arguments only--- are, as in
the previous model,
just products of the coefficients $\alpha_I,\gamma_I$ in
(\ref{sho:rep}). Analogous to (\ref{sum:ocond}), the coefficients
$\kappa_\pm$ satisfy
 \be   \label{dif:ocond}
  \kappa_+(j) \kappa_-(j)= j^2-\delta^2.
 \ee
Since only the $\kappa_\pm$ are relevant to the observable algebra, we
can view (\ref{dif:ocond}) as a condition on $\kappa_\pm$.
With this condition, the CCRs (\ref{dif:occr}) are also identically
satisfied.

Let us consider, as before, the completeness of the set of observables.
The classical analogs of $\{\hJ_z,\hJ_\pm\}$ are the functions
 \be  \label{dif:obs}
  J_z=\half(z_1\bar{z}_1+z_2\bar{z}_2),\quad J_+=z_1z_2,\quad
       J_-=\bar{z}_1\bar{z}_2
 \ee
on \ps. (Note that there is an ambiguity in the correspondence between
the operator $\hJ_z$ and the classical function $J_z$. This ambiguity
is resolved by requiring the Poisson bracket algebra between the
classical functions to be the Lie algebra of $SO(2,1)$.)  As in the
previous model, one can easily check that the set $(J_+,J_-)$ is by
itself {\em complete}; $J_z$ is included in order that the set of
observables is closed under the Poisson bracket. Again, there is an
algebraic relation between the overcomplete set of generators of \cAp\
which fixes the value of a Casimir invariant of \cAp. Using the
definitions of $\hJ_\pm$ and the commutation relations
(\ref{sho:ccr}), one finds that
 \be\label{dif:oacr}
  \hJ^2:=-\hJ_z^2+\half[\hJ_+,\hJ_-]_+ \equiv\tfrac14-\delta^2 \ .
 \ee
Equivalently, the algebraic identity can be expressed as
 $$\hJ_+\hJ_-=(\hJ_z-\half)^2-\delta^2, \eqno(\hbox{\ref{dif:oacr}}'),$$
in which form it is clear that the condition on the operators is
automatically satisfied due to (\ref{dif:ocond}).

The last step in the program is to select an inner product by
requiring that the $\star$-relations on \cAp\ become Hermitian
adjointness relations on the resulting Hilbert space.
As for the previous model, from the
expressions for the physical operators in terms of the elementary
quantum operators $\ha_I,\hc_I$ and the $\star$-relation
(\ref{sho:elemstar}), one obtains the $\star$-relation induced on
\cAp:
 \be  \label{dif:o*}
  \hJ_+^\star=\hJ_- \quad\hbox{and}\quad \hJ_z^\star=\hJ_z.
 \ee
The Hermiticity
condition on $\hJ_z$ requires that its eigenvalues $j$ must be real,
and its eigenkets orthogonal to each other. As in the previous model,
the representation (\ref{dif:orep}) is reducible: the physical
operators either leave the value of $j$ unchanged, or change it by an
{\it integer}.  Thus, the fractional part of $j$ ---denoted by
$\epsilon={\rm frac}(j)$--- is invariant under the action of
$\hJ_z,\hJ_\pm$. Consider ${\cal V}_{phy}^\epsilon$, the vector space
of states with the same fixed value of $\epsilon$. Each ${\cal
V}_{phy}^\epsilon$ carries an irreducible representation of the
$SO(2,1)$ Lie algebra (\ref{dif:occr}); however, the
$\star$-relations on the algebra have not all been imposed.

Let us now return to the Hermiticity conditions. Let
$j=n+\epsilon$, $n=\cdots -~2,-1, 0,1,2\cdots$.  Each ${\cal
V}_{phy}^\epsilon$ has a countable basis, labelled by $n$, the integer
part of $j$, and it is on these {\it irreducible} representations that
one implements the Hermiticity conditions on \cAp.  Note that at this
stage it appears that we have a 1-parameter family of ambiguities in
quantization of the system, labelled by the parameter $\epsilon\in
[0,1)$.

Henceforth, for definiteness, let us consider a representation with a fixed
value of $\epsilon$. The Hermiticity of $\hJ_z$ implies that on ${\cal
V}_{phy}^\epsilon$ there exists an inner product in which the above
basis is orthogonal; without any loss of generality, we can choose it
to be orthonormal. Hence the inner product can be chosen to be:
 \be   \label{dif:ip}
  \IP{j'=n'+\epsilon,\delta}{j=n+\epsilon,\delta}=\delta_{n',n},
 \ee
where both states on the left have the same fractional part of $j$.
As in the case of the energy sum model, it is only because
we implement the Hermiticity conditions on an {\em irreducible} sector
---with a countable basis--- that we can postulate a Kronecker-$\delta$
normalization on ${\cal V}_{phy}^\epsilon$. On \cVp, we would be led
to a Dirac-$\delta$ normalization.

Now, the first of the $\star$-relations (\ref{dif:o*}) implies that
$\kappa_+(j)=\ovr{\kappa_-(j)}$. Substituting this in
(\ref{dif:ocond}), the condition on the undetermined coefficients,
yields
 \be\label{dif:ocond2}
  |\kappa_+(j)|^2= j^2-\delta^2.
 \ee
We can use the freedom in the phase of the kets $\ket{j,\delta}$ to
make $\kappa_+(j)$ real for all $j$, and solve (\ref{dif:ocond2}).
Then we have
 \bea \label{dif:fixrep}
  \hJ_z\ket{j,\delta} &=& (j+\half)\ket{j,\delta}, \cr
  \hJ_+\ket{j,\delta} &=& \sqrt{(j+1)^2-\delta^2}\>\ket{j+1,\delta} \cr
  \hbox{and}\quad\hJ_-\ket{j,\delta} &=&
       \sqrt{j^2-\delta^2}\>\ket{j-1,\delta}.
 \eea
For solutions to exist, we require that physical states satisfy
 \be \label{dif:condsol}
  j^2\ge\delta^2.
 \ee
However, we cannot simply begin with any state satisfying
$j^2\ge\delta^2$ and hope to obtain
a genuine representation of the observable algebra.
Consider for example a state $\ket{j,\delta}$
with arbitrary $j\ge|\delta|$. In the simplest
cases (see the remark at the end of this section), using $\hJ_-$
repeatedly, one can lower $j$ until the condition $j^2\ge\delta^2$ is
violated, unless there exists a state $\ket{j_0,\delta}$ annihilated
by $\hJ_-$. From (\ref{dif:fixrep}), we see that this will occur if
and only if
$j_0=\pm|\delta|$.  Acting with $\hJ_+$ repeatedly on the ``ground''
state $\ket{j_0=|\delta|,\delta}$, we see that an allowed
representation consists
of states labelled by
 \be \label{dif:j+}
  j=|\delta|+n,\quad n=0,1,2\cdots .
 \ee
This corresponds to a representation with a fixed value
$\epsilon=\rmfrac(|\delta|)$.

Similarly, starting with arbitrary $j\le -|\delta|$ one can use
$\hJ_+$ repeatedly to raise $j$ until (\ref{dif:condsol}) is violated,
unless there exists a ``top'' state, $\ket{j_0,\delta}$ which is
annihilated by $\hJ_+$. This would happen if
$j_0=-|\delta|-1$. One obtains a representation inequivalent to the
previous one,
 \be   \label{dif:j-}
   j=-|\delta|-1-n,\quad n=0,1,2\cdots ,
 \ee
so that $\epsilon=1-\rmfrac(|\delta|)$. Thus for each value of the
energy difference $\delta$
one obtains the two representations (\ref{dif:j+}) and (\ref{dif:j-})
of the algebra of observables.

As in the quantization of the energy sum model, the coefficients
$\alpha_I, \gamma_I$ in (\ref{sho:rep}) are left undetermined, and we
have {\it not} obtained an inner product on the original
representation space. However, (\ref{dif:ip}) provides us with an
inner product on \cVp. Note that before imposing the reality
conditions, there was considerable ambiguity in the choice of
representation of the observable algebra. In the first place, there
are representations in which $j$ is complex-valued; next, there is
ambiguity in the choice of the coefficients $\kappa_\pm$ satisfying
(\ref{dif:ocond}). Further, for fixed choices of $\kappa_\pm$, there
is a 1-parameter family of irreducible representations of the
observable algebra. Imposing the $\star$-relations had four
consequences: {\em i}) We were restricted to representations with real
valued $j$; {\em ii}) we found an inner product on physical states;
{\em iii}) we found unique $\lambda_\pm$ satisfying (\ref{dif:ocond});
and, {\em iv}) in the simplest case considered above, for each choice
of $\delta$ we are left with only two irreducible representations,
which can be distinguished by the value of $\epsilon$.

In contrast to the energy sum model, where one finds representations
only for integer $E$, in this model there is no such constraint on
$\delta$. This difference between the two models is related to the
fact that in the energy sum model the reduced \ps\ is compact ($S^2$)
whereas for the energy difference model the reduced \ps\ is
non-compact (${\rm I}\!{\rm R}^2$). In quantum theory, this shows up
as a difference between the conditions on the parameters labelling
physical states: In the energy sum model $|m|$ is bounded, whereas in
the energy difference model, $|j|$ is bounded only from
below. Finally, whereas in the energy sum model there is a {\em
unique} representation for each integer $E$, in this model we have
{\em two} representations of the physical observable algebra for all
real $\delta$. Without further physical input, both are admissible
quantum theories.
\medskip

\noindent{\sl Other Representations:}

The representations we considered above are ``generic.'' There are, in
addition, some exceptional cases. From (\ref{dif:fixrep}) we see that
a ``ground'' state $\ket{j_0,\delta}$ is annihilated by $\hJ_-$ if and
only if $j_0=\pm|\delta|$. To construct the representation
(\ref{dif:j+}) we acted repeatedly with $\hJ_+$ on {\em one} choice of
ground state, namely $\ket{j_0=+|\delta|,\delta}$. Let us attempt to
construct a representation by acting repeatedly with $\hJ_+$ on the
other possible choice of ``ground state'':
$\ket{j_0=-|\delta|,\delta}$. Then the allowed states are labelled by
$j=-|\delta| + n$. However, generically, this procedure does not lead
to viable representations: in particular, the ``first excited state''
$\ket{j=-|\delta|+1,
\delta}$ violates (\ref{dif:condsol}) unless $|\delta|\le\half$.
Similarly, the alternate choice of ``top state'' $\ket{j_0=
+|\delta|-1}$ ---i.e. a state annihilated by $\hJ_+$--- does
not yield a new representation unless $|\delta|\le\half$. However, if
$0<|\delta|\le\half$ we {\em do} have the additional representations
 \be \label{dif:j+'}
   j=-|\delta|+n,\quad n=0,1,2\cdots\qquad 0<|\delta|\le\half\ ,
 \ee
for which $\epsilon=1-\rmfrac(|\delta|)$; and,
 \be \label{dif:j-'}
  j=|\delta|-1-n,\quad n=0,1,2\cdots\qquad 0<|\delta|\le\half\ ,
 \ee
with $\epsilon=\rmfrac(|\delta|)$.

In fact, as Louko has pointed out \cite{jl:pap}, for $|\delta|
\in[0,\half)$, the above representations are only special cases. There
is a whole slew of representations, one for each choice of
$\epsilon\in[|\delta|,\half]$ or $\epsilon\in[-\half,-|\delta|]$. In
each such representation, the basis states are labelled by {\em all}
integers:
 \be\label{dif:eprep}
  j=\epsilon + n, \quad n=\cdots -2,-1,0,1,2 \cdots \qquad
    |\delta|\le|\epsilon| \le \half,
 \ee
not just the non-negative ones as in (\ref{dif:j+}--\ref{dif:j-'}).
These representations do not possess either a ``ground state'' or a
``top state''.  Note also that of the new representations we have just
constructed, in all but the representation (\ref{dif:j+'}), $j$ is
unbounded below.

\subsection{Remarks}
The quantum theory of the ``energy difference'' model has a number of
interesting features which we can now discuss.
\medskip

\noindent{\sl Inner product on \cV:}

In the energy difference model, we imposed the Hermiticity conditions
only on physical states. We could of course have imposed them already
on the elementary operators, prior to solving the quantum constraint
equation and obtained a ``kinematic Hilbert space.''  However,
generically, we would have found that none of the solutions to the
constraints are normalizable with respect to that inner product. To
see this, note that, if the elementary operators are represented by
Hermitian operators, then the number operators $\hN_1$ and $\hN_2$
would take on only integral values. Hence, if $\delta$ is {\it not} an
integer or half-integer, the only normalizable state in the kernel of
the constraint operator would be the zero state.  Thus, we would be
forced to conclude that \cVp\ is zero dimensional. The resulting
quantum theory is clearly incorrect since, in this case, the reduced
phase space is a 2-dimensional {\em non-compact} manifold; the system has
one ``true'' degree of freedom.  Thus, our strategy of holding off the
imposition of the $\star$-relations until {\em after} the physical
states are isolated is {\it essential} to obtain an acceptable quantum
theory.
\medskip

\noindent{\sl Energy:}

Recall that, in the case of two ordinary (i.e. unconstrained)
oscillators, the function $H(x_I, p_I) := \half(x_1^2 + p_1^2 + x_2^2
+ p_2^2)$ is the total energy.  Let us therefore refer to the
corresponding operator $\hat{H}=2\hJ_z$ as the Hamiltonian (although
it may have nothing to do with the actual dynamics of the constrained
system). Now, in the classical theory, the total energy is
non-negative, i.e.\  $H\ge0$. In the quantum theory however, we have
obtained representations (the ones other than (\ref{dif:j+},
\ref{dif:j+'})) of \cAp\ in which the corresponding operator $\hat{H}$
is {\it unbounded below} (with eigenvalues $- 2|\delta|-2n-1,\>
n=0,1,2\cdots, \delta\in \real$; $2|\delta|-2n-1, n=0,1\cdots ,
|\delta|\le \half$, or, $2\epsilon +2 n +1,\> n=\cdots -1,0,1\cdots$).
Since $\hat{H}$ is an {\it elementary physical observable}, it is not
of the form $\hat{O}^\star \hat{O}$ for any $\hat{O}\in$\cAp; there is
nothing to ensure its positivity.  Thus, without additional physical
input, we can not rule out these representations. However, in the
representations corresponding to (\ref{dif:j-}) and (\ref{dif:j-'}),
there are {\it no} states with positive energy eigenvalues. Therefore,
by requiring in addition that we obtain an acceptable classical limit,
we can rule out the representations corresponding to $m<0$.

Even if we restrict ourselves to the positive energy representations,
the spectrum of the energy operator has a feature that is at first
unexpected. Recall that, in the quantum theory of two oscillators, the
eigenvalues of the total Hamiltonian can take only integral values.
In the present case, on the other hand, we found that $\hat{H}=
2\hat{J}_z$ whence its eigenvalues, ($2n+2|\delta|+1$), are in general
non-integral. How does this result come about? The answer is that it
is forced on us by the constraint. In the usual quantum theory of an
oscillator, it is the requirement of positivity of the energy and the
existence of an annihilation operator that forces the energy
eigenvalues to be half-integral. Because the quantum constraint is
already satisfied by the physical states, once the energy of the
``lower'' oscillator is positive, the constraint guarantees that the
energy of the ``higher'' oscillator will also be positive; this is no
longer an independent requirement! Hence the energy of the ``higher''
oscillator is not forced to be half-integer, whence the total energy
is also not subject to that requirement.
\medskip

\noindent{\sl Overcomplete algebra of observables:}

In both the ``energy sum'' and the ``energy difference'' models, one
can construct the reduced phase space $\hat\Gamma$ in a
straightforward manner \cite{thesis}. It turns out, however, that
$\hat\Gamma$ does not naturally inherit a cotangent bundle structure.
For the energy sum model, $\hat\Gamma$ is $S^2$, which is compact.
For the energy difference model, while $\hat\Gamma$ is topologically
${\rm I}\!{\rm R}^2$, the symplectic structure is not the obvious one;
it is more natural to think of $\hat\Gamma$ as the positive mass-shell
in a 3-dimensional Minkowski space \cite{thesis}. A related
interesting feature is that, although we began with algebraically
independent elementary variables (and did not therefore have to impose
the ACRs), the algebra of {\it physical} observables is {\it
over}complete. This is particularly striking in the energy difference
model where the topology of the reduced \ps\ is trivial. In both
examples, it is the requirement that the set of generators of \cAp\ be
closed under the commutator Lie bracket that forces one to include an
``extra'' element in the set of generators of \cAp. Thus, there is an
algebraic relation on the physical observables, the quantum version of
which is satisfied on the physical states.
\medskip

\noindent{\sl Discrete symmetries:}

Recall that prior to imposing the Hermiticity conditions, for a fixed
choice of $\kappa_\pm$ satisfying (\ref{dif:ocond}), we had a
1-parameter family, labelled by $\epsilon$, of irreducible
representations of the algebra of physical observables. It is natural
to ask if the irreducible sectors ${\cal V}_{phy}^\epsilon$ are the
eigenspaces of an operator corresponding to some (discrete) symmetry.

Let us begin with the classical theory. The constraint function
$\half(z_1\bar{z}_1-z_2\bar{z}_2)-\delta$ of (\ref{dif:zcon}) as well
as the physical observables $\{\half(z_1\bar{z}_1+z_2\bar{z}_2),
z_1z_2,\bar{z}_1\bar{z}_2\}$ of (\ref{dif:obs}), are all invariant
under the {\it discrete} ``parity'' map $(z_I)\longmapsto {\bf
P}(z_I):=(-z_I)$.  Hence, although the set of observables
(\ref{dif:obs}) is locally complete on the constraint surface
$\bar\Gamma$ (in the sense that their gradients span the cotangent
space at each point of $\bar\Gamma$) they fail to capture certain
global information about the constraint surface since they can not
distinguish between the point labelled by $z_I$ and that labelled by
$-z_I$. Let us now consider the quantum theory. In the Bargmann-type
representation, this discrete symmetry corresponds to the operation
$\psi(z_I)\mapsto\psi(-z_I)$. Since $\IP{z_1, z_2}{j,m} = z_1^{j+m}
z_2^{j-m}$, it follows that in the $\ket{j,m}$ representation,
the action of the parity operator is given by
 \be \label{dif:par}
  \ket{j,m} \longmapsto\hat{\bf P}\ket{j,m}:=(-1)^{2j}\ket{j,m},
 \ee
where, to evaluate the right hand side we will take the principal
value, namely, $(-1)^{2j}=\exp(i2\pi\epsilon)$, and as before
$\epsilon=\rmfrac(j)$ is the fractional part of $j$. Next, recall that
since the physical operators change $j$ only in integral steps they do
not affect the fractional part $\epsilon$.  Consequently, \cVp\ is
reducible, and {\em each eigenspace ${\cal V}_{phy}^\epsilon$ of the
parity operator provides an irreducible representation of the algebra
\cAp}. Thus, there is indeed a superselection.

The parity operator has an unexpected feature in quantum theory.
Classically, the parity transformation ${\bf P}(z_I)=(-z_I)$ satisfies
 \be  \label{sum:feat1}
  {\bf P}^2=1.
 \ee
irrespective of the precise value of $\delta\in \real$. In quantum
theory, on the other hand, the eigenvalues of $\hat{\bf P}$ are given
by $\exp(i2\pi\epsilon)$ (see (\ref{dif:par})), where in the physical
representation, $\epsilon =\rmfrac(|\delta|)$. Hence, on physical
states, $\hat{\bf P}^2=\exp(i4\pi\rmfrac(|\delta|)$. Thus, we recover
the classical behavior only if $\delta$ is an integer or half integer.
For other values, the classical behavior of the parity symmetry can
not be recovered. Within quantum mechanics, however, there seems to be
{\it no} compelling reason to restrict ourselves to states with
eigenvalues $\pm 1$ of $\hat{\bf P}$. On sectors with other
eigenvalues, we have a situation that is rather similar to the one
encountered in systems of identical particles in 2 space dimensions
where the use of eigenvalues other than $\pm 1$ for the parity (or the
permutation) operator leads to the interesting quantum phenomena of
fractional statistics.
\medskip

{\sl FRW universe with conformally coupled scalar field:}

As we remarked earlier, the energy difference model arises in quantum
cosmology. In the context of the algebraic program, this fact is
mainly a mathematical curiosity since the program is not equipped to
deal with the difficult interpretational issues faced by quantum
cosmology. Nonetheless, it is useful to regard this minisuperspace as
a toy model for full general relativity and see that the program does
lead to mathematically complete quantum descriptions without, e.g.,
having to first solve the problem of time.

Let us begin by indicating how the cosmological model can be reduced
to the energy difference model.  The Ricci scalar for the closed
($k=+1$) Friedman-Robertson-Walker universe is
 \be
  R= \frac6{a^2} \left( a\frac{\d^2a}{\d\tau^2} + \left(\frac{\d
    a}{\d\tau}\right)^2 + 1\right),
 \ee
where $a$ is the scale-factor of the universe and $\tau$ is the proper
time.  Hence, the gravitational part of the Lagrangian (up to a factor
of $\frac{4\pi}{3}$, and after an integration by parts) is:
 \be
  {\cal L}_G=\frac6{G} \left(-a\left(\frac{\d a}{\d\tau}\right)^2+a\right),
 \ee
where $G$ is the gravitational constant, and the action is $S=\int
\rd\tau {\cal L}$. The Lagrangian for the homogeneous, conformally
coupled ($\xi=\frac16$, massless) scalar field is
 \be
  {\cal L}_{KG}=8\pi \left( a^3\left(\frac{\d\phi}{\d\tau}\right)^2
   + a\left(\frac{\d a}{\d\tau}\right)^2\phi^2
   + 2a^2\phi\frac{\d a}{\d\tau}\cdot\frac{\d\phi}{\d\tau}
   - a\phi^2 \right)\ .
 \ee
Introduce a reparametrization of the time, $\d t=\d\tau/N$, where $N$
is the lapse.  Let $\dot{\hphantom{a}}\equiv(\d/\d t)$. Then the total
Lagrangian is
 \be \label{frw:lag}
  {\cal L}= -\frac{6}{GN}a\dot{a}^2 + \frac{6Na}{G}
   + \frac{8\pi a}{N}\dot{(a\phi)}^2 - \frac{8\pi N}{a}(a\phi)^2.
 \ee

Define the variables
 \bea  \label{frw:var}
             x_1 &:=& \sqrt{\frac{12}{G}}\,a \\
             x_2 &:=& \sqrt{\frac{16\pi}{G}}\,a\phi.
 \eea
Now the Lagrangian takes the form
 \be
  {\cal L}= -\frac{1}{2N}a\dot{x}_1^2 + \frac{1}{2N}a\dot{x}_2^2
    + \frac{6Na}{G} - \frac{N}{2a}x_2^2.
 \ee
Performing the Legendre transform, we find the canonical momenta:
 \bea
  p_1 &:=& -\sqrt{\dfrac{12}{G}} \dfrac{a\,\dot{a}}{N} \\ \hbox{and}
  p_2 &:=& \tfrac{4}{\sqrt\pi N}\,a\,\dot{(a\phi)}.
 \eea
Since we are in a spatially compact situation, the Hamiltonian is
constrained to vanish. Indeed, it has the form $H=\frac{2N}{x_1}C$,
where the scalar constraint $C$ is given by
 \be \label{frw:con}
  C={1\over 4}(p_1^2+x_1^2-p_2^2-x_2^2)\approx 0.
 \ee
We see that it is exactly of the form of (\ref{dif:zcon}), with
$\delta=0$.

Note, however, that there is a nonholonomic constraint, $a\ge0$. One
consistent approach to handle this would be to consider the physical
scale factor to be defined by $a:= \frac{|x_1|} {\sqrt{12}}$, on the
\ps\ defined by $(x_I,p_I)$.  The solutions then describe a periodic,
bouncing universe. One can use the representations obtained in
section 4.2 as quantum descriptions of this model%
\footnote{This cosmological model has been quantized elsewhere, see e.g.\
\cite{gmm:kg}, in which the quantum theory corresponds to the
representation (\ref{dif:j+}) and does not include any of the
representations (\ref{dif:eprep}).}.

\mysection{Constrained rotor model}
\newcommand\ahpre{(\sin\theta)^{-\half}}

In this section we will quantize a model which mimics some of the
features of \gr. It was introduced by Ashtekar and Horowitz
\cite{aa:gh} and was believed to display certain unexpected behaviour
in the quantum theory. Since this behaviour arose from precisely those
features of this model that it shares with \gr, there was some concern
that similar surprises might occur in a quantum theory of gravity. In
\cite{aa:gh}, however, there were no guidelines to select the inner
product on the physical states. By using the algebraic program, we can
now select the ``correct'' inner-product and analyse the issues raised
in \cite{aa:gh} within the resulting quantum representation. We will
find that the unexpected features are in fact absent; they arose
because of the use of a physically incorrect inner-product. The manner
in which the correct inner product avoids the problems is quite subtle
and, without the reality conditions to guide us, it would have been
difficult to argue that this is the appropriate inner product to use.
Thus, the example illustrates the power of the algebraic approach in
addressing concrete physical issues.

\subsection{Motivation}
Let us recall certain features of \gr\ in the geometrodynamical
variables.  In the Arnowitt-Deser-Misner (ADM) formulation \cite{adm},
the basic \ps\ variables are the 3-metric and its canonically
conjugate momentum. The constraint surface%
\footnote{In these motivating remarks, the diffeomorphism constraint
 of \gr\ plays no role and is therefore ignored.}
is specified by the vanishing of the scalar constraint function. The
scalar constraint is the sum of two terms: a kinetic term ---quadratic
in the momenta--- the coefficient of which defines a ``supermetric''
on the configuration space; and a potential term ---proportional to
the 3-dimensional Ricci scalar--- which depends only on the
configuration variables. Due to the complicated form of the
constraint, the geometry of the constraint surface, $\Gbar$, and the
structure of the reduced \ps\ of \gr\ are still not fully
understood. There are, however, many features which are well known.
Of interest to us are the following:
\begin{itemize}
\item First, the constraint surface defines a ``classically forbidden''
region in the configuration space. More precisely, the image in the
configuration space of the constraint surface $\Gbar$ (under the
natural projection map) is a {\it proper subset\ } of the
configuration space $\cC$.
\item Second, in the asymptotically flat
case, the Hamiltonian is not constrained to vanish. On $\Gbar$, the
Hamiltonian reduces to a surface integral at spatial infinity, called
the ADM energy. The ADM energy depends only on the 3-metric and its
spatial derivatives, and not on the canonically conjugate momenta.
\item Finally, the positive energy theorems of
classical \gr\ state that on the allowed regions of $\cC$ defined by
the projection of the constraint surface, the ADM energy is positive.
In the ``forbidden'' regions, where the constraint cannot be
satisfied, the ADM energy can be negative.
\end{itemize}

Ashtekar and Horowitz constructed a finite dimensional model which
mimics the above features of \gr. Consider a particle on a (unit)
2-sphere in 3-dimensional Euclidean space%
\footnote{The inclusion or exclusion of the radial degree of freedom
 plays no significant role.},
subject to the constraint
 \be \label{ah:con} C\equiv p_\theta^2-R(\phi)=0,  \ee
where $(\theta,\phi)$ are the usual spherical coordinates.  The
``potential'', $R(\phi)$, is a smooth function, which is not
everywhere positive. As in \gr, the constraint surface $\Gbar$
projects down to a proper subset of the configuration space $\cC$: the
{\em classically allowed region} $\Cbar$ corresponds to those sectors
where (\ref{ah:con}) has solutions, i.e., where $R(\phi) \ge0$.  Now
introduce a Hamiltonian via
 \be\label{ah:ham}
   H=C+E(\phi), \quad \hbox{with} \quad  E(\phi)\cdot R(\phi)\ge0; \ee
we assume that $E$ is bounded.  On the constraint surface $C=0$, the
Hamiltonian reduces to the ``ADM energy'' $E(\phi)$, and depends only
on the configuration variable $\phi$. Since $E(\phi)$ is positive in
the classically allowed regions, where $R(\phi)$ is positive, this
function satisfies a classical positive energy theorem, as does the
Hamiltonian in \gr. We will henceforth refer to this
model as the {\em constrained rotor model}.

Now consider the Dirac quantum theory of this model, say in the
Schr\"odinger representation, where states are functions of the
configuration variables. Ashtekar and Horowitz raised the following
question: In the Dirac quantum theory, do there exist physical states
that penetrate the classically forbidden region ($R<0$)?  Let us
suppose that such states do exist.  Now, physical states are solutions
to the quantum constraint equation, and on these states the
Hamiltonian acts via a multiplication by $E(\phi)$. (Note that ---as
in \gr--- in this model, the ``ADM energy'' $E$ is a function only of
the configuration variables.) In this model $E$ is negative in the
{\em classically forbidden region} $\tcbar=\cC - \Cbar$. Thus, on some
physical states which penetrate sufficiently into $\tcbar$, the energy
will be negative. Due to the close analogy with \gr, if such behaviour
occurs in this model it would indicate that similar tunnelling can
occur in quantum gravity as well.

We will re-analyse this problem in the context of the
algebraic quantization program.

\subsection{Dirac quantization}

Let $(\theta, \phi)$ denote the usual spherical coordinates on $S^2$,
and let $(p_\theta, p_\phi)$ be the canonically conjugate momentum
operators. We choose the set \cS\ of elementary variables to consist of all
functions $f(\theta,\phi)$ on $S^2$ and the momentum variables
$p_\theta,p_\phi$. The $\star$-algebra \cAs\ is straightforward to
construct.  An obvious choice for the vector space representation is
provided by the availability of a configuration space. Here, the
states are (complex-valued) functions on $S^2$, $\Psi=\Psi(\theta,
\phi)$, and the elementary operators are represented by
 \bea \label{ah:rep}
 \hat{f}\circ\Psi(\theta,\phi) &=&
 f(\theta,\phi)\cdot\Psi(\theta,\phi)     \cr
 \ptthat\circ\Psi(\theta,\phi)&=&\dfrac\hbar{i}\dfrac1{\sqrt{\sin\theta}}
 \dfrac\d{\d\theta}\left(\sqrt{\sin\theta}\,\Psi(\theta,\phi)\right)
 \equiv\dfrac\hbar{i}\left(\dfrac\d{\d\theta}
 + \frac12\cot\theta\right)\Psi(\theta,\phi) \cr
 {\hbox{and}}
   \quad \pphihat\circ\Psi(\theta,\phi)&=&
 \dfrac\hbar{i}\dfrac\d{\d\phi}\Psi(\theta,\phi) .
 \eea
These operators satisfy the usual CCRs, and although the set of
elementary variables is overcomplete, all the resulting
anticommutation relations are identically satisfied in this
representation. Note that we do {\em not} require the Hermiticity of
the elementary operators, and since we have no inner product, we could
have left the familiar ``divergence'' term out of the representation
of the momentum operator $\ptthat$%
\footnote{In fact, as we discuss briefly in section 5.3,
 requiring that the momentum operators be Hermitian on all states (as
 in \cite{db:ah}) leads to an inadequate space of physical states.}.
We could define a kinder, gentler, representation by replacing
$\sqrt{\sin\theta}\Psi$ by $\Psi$; then, for example,
$\ptthat\circ\Psi=(\hbar/i)(\d/\d\theta)\Psi$ and the forms of the
constraint and the solutions are simplified considerably. However, we
will use the above representation (\ref{ah:rep}), as it is the one
chosen by Ashtekar and Horowitz%
\footnote{The vector field $\d/\d\theta$ is not globally defined on
 $S^2$. However, this is not essential to the points we illustrate here,
 since in particular the analysis could be repeated by replacing $S^2$ by
 a cylinder, where this subtlety is not encountered. Nonetheless, in
 order not to obscure the main point ---the power of the algebraic
 approach to remove ambiguities--- we will continue
 to use the original model discussed in the literature.}.
The two representations are equivalent (see \cite{thesis}). In the
representation (\ref{ah:rep}), the constraint equation satisfied by
physical states $\psi$ is
 \be \label{ah:qcon}
   \dfrac{\hbar^2}{\sqrt{\sin\theta}}\fr{\d^2}{\d\theta^2}
   \left(\sqrt{\sin\theta}
  \psi\right) + R(\phi)\psi=0.
 \ee
Although we assume for simplicity that $R(\phi)$ is smooth,
discontinuities in $R$ will not matter.

Since this is a quadratic differential equation, there are two linearly
independent sets of solutions
 \be  \label{ah:qsol}
   \psi_\pm=k_\pm(\phi)\cdot\ahpre\cdot
   \exp({\pm\fr{i}\hbar\sqrt{R}\theta})
 \ee
where $\sqrt{R}$ denotes the principal value: $\sqrt{R} =+\sqrt{R}$,
if $R>0$, and $\sqrt{R} =+i\sqrt{|R|}$ if $R<0$. The functions
$k_\pm(\phi)$ are arbitrary and have support everywhere in
configuration space, including the classically forbidden region
$\tcbar$. Let the linear vector space of physical states be denoted by
$\cV$. Then $\cV=\cV^+\oplus\cV^-$, where $\psi_\pm \in
\cV^\pm$ respectively, and we can write a general (physical) state as
 \be \label{ah:gsol} \psi=\left(\ba{c} \psi_+ \\ \psi_- \ea \right).
   \ee
Note that since we have not yet defined an inner product, this
representation of $\psi$ does {\em not} provide us with an
orthogonal decomposition of $\cV$.

Let us construct the Dirac observables. Since the constraint is first
class, there is one true degree of freedom, and thus we expect two
independent Dirac observables. From their representations, it is clear
that $\ptthat, \widehat{f(\phi)}$ commute with the constraint and
leave $\cV$ invariant, their only effect on the physical states is to
change the coefficients $k_\pm$ of the corresponding exponential
terms. However, on the constraint surface $p_\theta
=\pm\sqrt{R(\phi)}$, and thus $(p_\theta,f(\phi))$ is not a complete
set.  In order for the set to be complete, we need another observable.
Ashtekar and Horowitz introduced a classical Dirac observable, ${\rm
P}_\phi^\pm=p_\phi\mp\frac12 \frac{\theta R'}{\sqrt{R}}$. The set
$(p_\theta,f(\phi),{\rm P}_\phi)$ is now complete almost%
\footnote{It fails to be complete only at those points on
 the constraint surface where $R(\phi)=0$. This fact plays a minor role
 in the Dirac quantization and will therefore be ignored in what follows.
 It is, however, quite significant in the construction of the reduced \ps\
 quantum theory \cite{thesis}.}
everywhere, in the required sense.

The quantum operator corresponding to ${\rm{P}}^\pm_\phi$ is:
 \be  \label{ah:tp1}
  \Pphihat= \left(\ba{ll} \Pphihatp & 0 \\ 0 & \Pphihatm \ea \right),
  \quad
  \hbox{where}\quad \Pphihat{}^\pm=\pphihat \mp
   \frac12\wh{\left(\frac{R'}{\sqrt{R}}\theta\right)},
 \ee
where $'\equiv\d/\d\phi\,$ denotes the partial derivative \wrt\
$\phi$. To see that this is a physical operator, let us concentrate on
one component, acting on $\cV^+$.  Then,
 \be  \label{ah:tpyes}
  \Pphihatp\circ\psi_+=\frac\hbar{i}\, k'_+\, \ahpre\cdot
   \eip \> \in\> \cV^+.
 \ee
Similarly, $\Pphihatm\circ\psi_-\in\cV^-$. Hence $\Pphihat$ as defined
above is a physical operator. For future reference, note that
 \be  \label{ah:tpno}
  \Pphihatm\circ\psi_+=\left(\frac\hbar{i}k'_+ \>+\>
2\left(\sqrt{R}\right)'\, \theta k_+
       \right)\cdot\ahpre\cdot\eip \> \not\in\>\cV,
 \ee
is not a physical state, since the second term in the coefficient in
the bracket is no longer a function only of $\phi$.

We now have to find the $\star$-relations on the algebra of
observables.  Clearly, $\ptthat^\star=\ptthat$, $\widehat{f(\phi)}^\star=
\widehat{f(\phi)}$ and are physical operators. In order to analyse the
$\star$-relation on $\Pphihat$, let us further decompose $\cV^+$ into
the set of states $\bcV^+$ with support only on the classically
allowed region $\Cbar$ and the set of states $\tcV^+$ with support
entirely in the classically forbidden region $\tcbar$. Now, on the
sector of states $\bcV^+$, $(\wh{\sqrt{R}})^\star=\wh{\sqrt{R}}$ and
hence
 \be   \label{ah:tpystar}
  (\Pphihatp)^\star=\left(\pphihat
   -\frac12\wh{\left(\frac{R'}{\sqrt{R}}\theta\right)}\right)^\star
   =\pphihat-\frac12\wh{\left(\frac{R'}{\sqrt{R}}\theta\right)}
   =\Pphihatp
 \ee
is a physical operator on $\bcV^+$. However, on the ``forbidden sector''
$\tcV^+$, $(\wh{\sqrt{R}})^\star=-\wh{\sqrt{R}}$ and hence
 \be  \label{ah:tpnostar}
  (\Pphihatp)^\star=\left(\pphihat
   -\frac12\wh{\left(\frac{R'}{\sqrt{R}}\theta\right)}\right)^\star
   =\pphihat+\frac12\wh{\left(\frac{R'}{\sqrt{R}}\theta\right)}
   =\Pphihatm,
 \ee
which, as we see from (\ref{ah:tpno}), is {\em not} a physical
operator.  In the matrix notation, since $(\Pphihat{}^\pm)^\star
=\Pphihat{}^\pm$ on the sector of `` classically allowed'' states
$\bcV$, $\,\Pphihat^\star=\Pphihat$ is a physical operator. On the
other hand, since on the sector of ``classically forbidden states''
$\tcV$,
 \be   \label{ah:tpnostar2}
  (\Pphihat)^\star=
    \left(\ba{cc} (\Pphihat^+)^\star & 0 \\ 0 & (\Pphihat^-)^\star \ea\right)
    = \left(\ba{cc} \Pphihat^- & 0 \\ 0 & \Pphihat^+ \ea\right)\ ,
 \ee
it does not leave the space of physical states invariant.

Clearly something peculiar is happening here. We have a complete set
of physical states which carry a representation of an (almost)
complete algebra $\cAp\subset\cA$ of physical operators generated by
$\{\ptthat, \widehat{f(\phi)}, \Pphihat\}$.  Further, the
$\star$-involution on \cA\ has a well-defined action on
\cAp\, and induces a map from \cAp\ into \cA. What fails however, is
that the induced $\star$ is {\em not an involution} on \cAp; its
action on one of the generators of $\cAp$, namely $\Pphihat$, takes it
out of \cAp. (This can be understood in terms of the Hamiltonian
vector field of the classical observable $\Pphi$, see \cite{thesis}.)
The algebra of physical observables, \cAp, does not admit a
$\star$-involution induced from \cAs. Thus there is no sensible way to
formulate the Hermiticity conditions on physical operators in terms of
an inner product on physical states.

On the face of it, due to the above mathematical inconsistency, i.e.\
the lack of a $\star$-involution on \cAp, one cannot proceed with the
quantization program. Since the difficulty arises due to the sector of
``forbidden'' states, one way out would be to discard them on
mathematical grounds. However, this is somewhat unsatisfactory since
there appears to be no compelling physical reason to do so. We would
be ruling out, {\it by fiat}, precisely the ``tunnelling'' states
whose existence is the issue under investigation.

An alternative approach would be to try and implement the
$\star$-relations on the other physical observables, and then see if
it leads to a mathematically and physically sensible framework. As we
will see in detail, we will find that the resulting measure is such
that the forbidden region $\tcbar$ is a set of measure zero, and
$\Pphihat{}^\star$ is an observable. Thus, it is not that the
solutions can not have support on the classically forbidden
regions. Rather, the forbidden region does not ``contribute'' because
it is simply a set of measure zero.

For the purposes of the analysis above, we had decomposed $\cV$ into
two linearly independent parts $\cV^+$ and $\cV^-$. Each of these
sectors carries an irreducible representation of $\cAp$, and one might be
tempted to consider an inner product in which $\cV^+$ and $\cV^-$ are
mutually {\em orthogonal}. This would be justified {\em if} we knew
that $\cV^\pm$ were the eigenspaces of some operator which is expected
to be Hermitian or unitary. In the absence of an obvious candidate for
such an operator, let us consider, as an ansatz, a general inner
product of the form
 \bea  \label{ah:ip1}
  \IP{\psi_1}{\psi_2}=    &\hphantom{+}
       {\displaystyle\int_{\scriptstyle\Cbar}}\> d\theta\wedge d\phi
   \quad (\mu_+\kopb\ktp + \mu_-\komb\ktm + \mu_{+-}\kopb\ktm
      + \ovr\mu_{+-}\komb\ktp) \cr
    & + {\displaystyle\int_{\scriptstyle\tcbar}}\> d\theta\wedge d\phi\quad
         (\tmu_+\kopb\ktp
      + \tmu_-\komb\ktm + \tmu_{+-}\kopb\ktm + \ovr\tmu_{+-}\komb\ktp)\, .
 \eea
The
$\mu=\mu(\theta,\phi)$ are arbitrary functions on $S^2$ and we have
absorbed an overall factor of $\sin^{-1}\theta$ into their
definitions.
Extra factors of $\exp(-\frac{2i}\hbar\sqrt{R}\theta)$,
$\exp(-\frac{2}\hbar\sqrt{|R|}\theta)$, and
$\exp(\frac{2}\hbar\sqrt{|R|}\theta)$ have been absorbed into $\mu_{+-}$,
$\tmu_+$ and $\tmu_-$ respectively%
\footnote{In terms of the matrix notation, the inner product corresponds to
 \be  \label{ah:ipmat}
  \IP{\psi_1}{\psi_2}= \ba{cc} (\>\kopb & \komb\>) \\
   & \ea \left(\ba{ll} \mu_+ & \mu_{+-} \\ \ovr{\mu}_{+-} & \mu_- \ea \right)
   \left(\ba{c}  \ktp \\ \ktm \ea\right)
 \ee
Suppose we write a general physical state as $\psi=\psi_+ + \psi_-$,
and postulate the seemingly natural inner product $\IP{\psi_1}{\psi_2}
=\int_{\Cbar} \mu \>\ovr{\psi_1}\,\psi_2 +\int_{\tcbar} \tmu
\>\ovr{\psi_1}\, \psi_2$. In the above notation, this corresponds to
choosing a matrix all of whose components are equal up to phase. A
short calculation, attempting to impose the Hermiticity conditions on
$\ptthat$, shows that there is in fact no such inner product.
Therefore one is forced to work with the more general form
(\ref{ah:ip1}) or (\ref{ah:ipmat}).}.
The inner product is positive definite \iff\ the ``diagonal'' measures
are positive and the ``off-diagonal'' measures satisfy
 \be  \label{ah:ip>0}
  |\mu_{+-}|<\sqrt{\mu_+\mu_-} \quad\hbox{and}\quad
           |\tmu_{+-}|<\sqrt{\tmu_+\tmu_-}.
 \ee

Now let us impose the Hermiticity conditions on the observables.
Clearly, all $\widehat{f(\phi)}$ are Hermitian.  Next, consider
$\ptthat$. We will carry-out a straightforward analysis which shows
that $\ptthat$ can be Hermitian only if $\tmu_+=\tmu_-=\tmu_{+-}=0$.

Since physical states are specified entirely by their coefficients
$k_\pm$, we can represent operators by their action on $k_\pm$.  In
particular,
\bea \label{ah:krep2} \ptthat\circ \left(\ba{c} k_+ \\ k_-
 \ea\right) = & \left(\ba{c} +\sqrt{R} k_+ \\
 -\sqrt{R} k_- \ea\right)\quad & \hbox{on}\quad \Cbar \cr = &
 \left(\ba{c} +i\sqrt{|R|} k_+ \\
 -i\sqrt{|R|} k_- \ea\right) \quad & \hbox{on}\quad \tcbar.
\eea
We want to find an inner product such that
$\bra{\psi_1}\ptthat\ket{\psi_2} =
\bra{\psi_1}\ptthat{}^\dagger\ket{\psi_2}$ for all
$\psi_1,\psi_2$. Consider states in $\cV_+$, i.e. $\kom=\ktm=0$. Then
\be \label{ah:iptp} \bra{\psi_1}\ptthat\ket{\psi_2} =
 \int_{\Cbar}\>\mu_+ \>\kopb\,\sqrt{R}\,\ktp
 +i\int_{\tcbar}\>\tmu_+\>\kopb\,\sqrt{|R|}\,\ktp.
\ee
On the other hand, by definition
\be \label{ah:iptpstar}
 \bra{\psi_1}\ptthat^\dagger\ket{\psi_2}
   :=\ovr{\bra{\psi_2}\ptthat\ket{\psi_1}}
   =   \int_{\Cbar}\>\mu_+  \>\kopb\,\sqrt{R}\,\ktp
     -i\int_{\tcbar}\>\tmu_+\>\kopb\,\sqrt{|R|}\,\ktp.
\ee
Clearly, $\ptthat$ can be Hermitian on $\cV_+$ only if $\tmu_+=0$.
Similarly, one can conclude that $\tmu_-=0$. Thus, both the diagonal
terms in the forbidden region vanish. But now, due to the condition
(\ref{ah:ip>0}), in the forbidden region the cross term also vanishes,
$\tmu_{+-}=0$. {\em In the inner product {\rm (\ref{ah:ip1})}, the
only terms that survive are the integrals over the classically allowed
region!} Thus, while physical solutions can indeed have support in the
classically forbidden region $\tcbar$, this support is of measure zero
in the physical inner product.

Let us return to the Hermiticity of $\ptthat$. Consider states
$\psi_1=\psi_{1+}$ and $\psi_2=\psi_{2-}$, i.e. $\kom=\ktp=0$. Then,
\be
 \bra{\psi_1} \ptthat \ket{\psi_2} = \int_{\Cbar}\> \mu_{+-}\,
 \kopb(-\sqrt{R}) \ktm.
\ee
However,
\be
 \bra{\psi_1} \ptthat^\dagger
 \ket{\psi_2} := \ovr{\bra{\psi_2} \ptthat \ket{\psi_1}} =
 \int_{\Cbar}\>\mu_{+-}\, \ktm(\sqrt{R})\kopb.
\ee
Equating the two, we find that $\ptthat$ is Hermitian if and only if
$\mu_{+-}=0$. Thus, it is the Hermiticity of a continuous operator
which implies that the subspaces $\cV_+$ and $\cV-$ are
orthogonal. (Since they are in fact orthogonal, clearly the system
admits another discrete symmetry, albeit a hidden one.)

Imposing the $\star$-relations on $\ptthat,\phihat$ as Hermiticity
conditions on the representation, we have reduced the form of the
physical inner product (\ref{ah:ip1}) to
 \be  \label{ah:ipred}
  \IP{\psi_1}{\psi_2}=\int_{\Cbar}\>(\mu_+\,\kopb\,\ktp
     + \mu_-\,\komb\,\ktm),
 \ee
where $\mu=\mu(\theta,\phi)$. Since the measure in the classically
forbidden region vanishes, as elements in the Hilbert space physical
states can be specified entirely by their support on $\Cbar$. In other
words, by restricting the support of physical states to the
classically allowed region $\Cbar$, in this problem we do not lose any
elements of the physical Hilbert space. Thus, as we saw earlier
(\ref{ah:tpyes}), on states with support only on $\Cbar$,
$\Pphihat^\star$ is a physical operator, in fact
$\Pphihat^\star=\Pphihat$. Now, since in terms of $k_\pm$, the action
of $\Pphihat$ on physical states is
 \be \Pphihat\circ k_\pm = \frac\hbar{i} \frac\d{\d\phi} k_\pm .  \ee
it is trivial to check that $\Pphihat$ is symmetric
if and only if $\d\mu_\pm/\d\phi =0$.

Thus, the measure depends only on $\theta$, and furthermore, this
dependence is not determined by any Hermiticity conditions.  Now,
since the coefficients $k_\pm$ do not depend on $\theta$ either, the
integral over $\theta$ can be performed trivially, and the inner
product is thus reduced to
 \be \label{ah:ipred2}
  \IP{\psi_1}{\psi_2}= \mu_+\int_{\ovr\phi}\> d\phi \, \kopb\ktp
                     + \mu_-\int_{\ovr\phi}\> d\phi \, \komb\ktm,
 \ee
where $\ovr\phi$ indicates that the integral is performed only over
classically allowed values of $\phi$, and $\mu_\pm$ are positive
numbers which are the results of the $\theta$-integration, or the
total measure on $\theta$.

Is there a criterion that will fix the relative weights of the two
terms? Recall that all the (continuous) physical observables we have
considered so far are diagonal in the representation (\ref{ah:gsol}),
i.e.\ their action leaves each of the physical subspaces $\cV^\pm$
invariant. In order to fix the relative weights of the two terms in
the inner product, we are led to look for a physical operator whose
action on physical states is not diagonal in the representation
(\ref{ah:gsol}); requiring it to be Hermitian or unitary would then
fix the inner product. It is natural to suppose that such an operator
corresponds to a discrete symmetry of the constraint.

An obvious symmetry is reflection in the $x$-$y$ plane,
$I_z:\theta\mapsto \pi-\theta$. In quantum theory, the corresponding
operator is represented by
$\hat{I}_z\circ\Psi(\theta,\phi)=\Psi(\pi-\theta,\phi)$. It is
manifestly a physical operator, and since classically $I_z^2=1$, it
should be both Hermitian and unitary, and its eigenspaces should be
orthogonal.  The even and odd physical eigenstates (with eigenvalues
$+1$ and $-1$ respectively) are of the form
 \bea \label{ah:ve}
  \psi_e &=\left(\ba{c} \hphantom{-} \frac12 k_e(\phi)
   \exp\left(-\frac{i\pi}{2\hbar}\sqrt{R}\right)\,\ahpre\cdot\eip \\
   \hphantom{-} \frac12
   k_e(\phi)\exp\left(+\frac{i\pi}{2\hbar}\sqrt{R}\right)\,
    \ahpre\cdot\eim
                   \ea\right) \\
      \label{ah:vo}
  \psi_o &=\left(\ba{c} \hphantom{-} \frac12 k_o(\phi)
     \exp\left(-\frac{i\pi}{2\hbar}\sqrt{R}\right)\,\ahpre\cdot\eip \\
    -\frac12 k_o(\phi) \exp\left(+\frac{i\pi}{2\hbar}\sqrt{R}
       \right)\,\ahpre\cdot\eim \ea\right),
 \eea
where $k_e$ and $k_o$ are arbitrary functions of $\phi$.  The
eigenspaces $\cV_e,\cV_o$ (\ref{ah:ve},\ref{ah:vo}) are orthogonal if
and only if $\mu_+=\mu_-$.  Thus we can choose $\mu_+=\mu_-=1$, and
the final form for the inner product is
 \be  \label{ah:ipfin}
  \IP{\psi_1}{\psi_2}= \int_{\ovr\phi}\> d\phi \>
       [\kopb\,\ktp + \komb\,\ktm].
 \ee

Is $\hat{I}_z$ ``super-selected'' in the sense that it commutes with
all other observables? If it were, then its eigenspaces would carry
irreducible representations of the algebra \cAp\ of (continuous) Dirac
operators. However, while it does commute with $\widehat{f(\phi)},
\Pphihat$, it does not commute with $\ptthat$ (In fact, they
anticommute, $[\ptthat,\hat{I}_z]_+=0$.) and hence $\ptthat$ is not
diagonal in the even/odd decomposition.

\subsection{Remarks}

Let us first briefly review the process by which we obtained a
complete quantum theory for this model. We chose a representation of
the elementary operators, on some vector space of complex valued
functions on the configuration space. In this representation, the
constraint equation is a second order partial differential equation,
which we solved explicitly. This set of solutions is ``large'' in the
sense that it includes the tunnelling solutions, which penetrate the
classically forbidden region.  Next, we constructed a set of
generators of \cAp, the algebra of physical observables. These
operators act on the ``large'' space of solutions and leave it
invariant. Then we attempted to induce the $\star$-involution on \cAp,
from \cA. The $\star$s of most of the generators of \cAp\ were also in
\cAp. However, the $\star$ (evaluated in \cA) of one of the physical
operators was no longer a physical operator itself.  Thus, we were
unable to induce on \cAp\ the structure of a $\star$-algebra.  This
appeared to be an impass\`e, in terms of constructing the physical
inner product via the prescription of the algebraic approach.

At this point we attempted to implement part of the $\star$-relations
as Hermiticity conditions on {\it some} of the physical operators.
Quite unexpectedly, these conditions led us to the conclusion that
mathematically distinct solutions to the constraint equation can lead
to the {\em same physical state}: the measure ignores the part of the
wavefunctions that corresponds to the tunnelling solutions. Finally,
the algebra of operators on the space of physical states ---which {\em
effectively} ignores the tunnelling solutions--- {\it does} admit a
$\star$-involution. Hence we were able to complete the quantization
program. Finally, let us consider the Hamiltonian (\ref{ah:ham}) in
the quantum theory we have constructed.  It is manifestly positive. On
physical states the Hamiltonian operator is
$\wh{H}\circ\psi=E(\phi)\cdot\psi$.  Since the states have
(measurable) support only in $\Cbar$, where $E(\phi)\ge0$,
$\langle\wh{H}\rangle\ge0$ for all physical states.

A key result of the Ashtekar-Horowitz quantization was that the
Dirac quantum theory does possess physical quantum states which tunnel
into the forbidden region $\tcbar$.  Further, some states had support
entirely in $\tcbar$! This led to the further conclusion that a large
number of physical quantum states possess negative energies: for
this model the classical positive energy theorem was violated in
quantum theory. These results were arrived at by using the ``obvious''
choice of measure: the Euclidean measure on ${\rm I}\!{\rm R}^3$. What
we have shown here is that the choice of the measure ---and hence of
the physical inner product--- is severely limited by the reality
conditions. The ``obvious'' choice is in fact incorrect.

A careful re-analysis of the problem has thus removed the exotic
features that were present in the previous quantization. While the
final result seems somewhat disappointing (in the sense that the
quantum theory does not display any exotic features), this example
serves to demonstrate the power of the quantization program of section
2. In retrospect our final result could have been obtained by a
number of other quantization procedures. However, in other schemes the
elimination of the spurious tunnelling states would have probably been
an {\it ad hoc} step.

In fact, based on an analysis of a similar model, which is perhaps
even more peculiar than the A-H model, Gotay \cite{mjg:ah} proposed
exactly such a requirement for quantum theory: {\it by fiat} the
measure is restricted to the classically allowed region. On the other
hand; in the algebraic approach to quantization, this result is {\em
derived from a general principle} (which is not expressly invented for
this example just to satisfy our classical intuition): real physical
operators should be Hermitian with respect to the physical inner
product.

In another re-analysis of this model, Boulware \cite{db:ah} assumed
the usual Euclidean inner product on the representation space, and
required that $p_\theta$ be a symmetric operator on the space of {\em
all} $L^2(S^2)$ states {\em before} solving the constraint equation
and isolating the physical states.  Then, physical states have support
only on the classically allowed region, and there is no tunnelling in
this theory.  As in the case of Gotay's analysis, the absence of
tunnelling is simply an input. Furthermore, now a new problem arises:
the resulting Hilbert space of physical states is only {\em finite
dimensional}, and therefore represents zero degrees of freedom!
However, the reduced \ps\ is a perfectly well-defined co-tangent
bundle over $S^1$ \cite{thesis}. Thus the quantum theory of
\cite{db:ah} does not capture all the physics in the model.

In the quantum theory we constructed here, since $\ptthat$ is an
observable, we too required it to be Hermitian, {\it but only on
physical states}.  Since on physical states $\ptthat=\sqrt{R(\phi)}$
one might wonder whether it is necessary to include $p_\theta$ in a
set of generators of the physical observable algebra. However, it is a
physical quantity, and one might wish to measure it. In the context
of the algebraic approach, the mathematically important facet of the
Hermiticity of $\ptthat$ is the following: as we saw in
(\ref{ah:tpnostar}), unless we impose the Hermiticity of $\ptthat$,
the rest of the formulation is {\em not mathematically well-defined}.

The quantization of a constrained system is inherently an ambiguous
process. One can only require that the final quantum theory is
``complete and consistent'' in the sense that one has a faithful
$\star$-representation of a suitably large algebra of observables and
that one recover the classical description in a suitable limit. To
achieve this end, one is justified in riding roughshod over the
initial stages of the road to quantization.  Thus, one can view the
quantization of the rigid rotor model in the following way: In the
beginning of this section, we exploited the freedom to define the
representation space and the operators, and were intentionally obscure
about the specification of \cV.  We then made some choice of
factor-ordering for the constraint and solved it on this space. On
physical states, we defined the actions of operators which formally
had vanishing commutators with the constraint. Then we imposed reality
conditions on these and found appropriate Hermitian physical
operators, completing the quantization program.

Finally, one can construct the reduced space quantum theory for this
model. The reduced \ps\ constructed in \cite{aa:gh}
appears to consist of two halves $\hat\Gamma_\pm$, each coordinatized
respectively by $(\bar\phi,\Pphi^\pm)$, where $\bar\phi$ indicates the
classically allowed values of $\phi$. The resulting reduced space
quantum theory corresponds to the quantization of a particle whose
\ps\ is the cotangent bundle over two disconnected intervals, and it is
relatively easy to see \cite{thesis} that it is unitarily equivalent
to the Dirac theory we have constructed above.  However, there is a
subtlety in the construction of the reduced \ps\ which arises due to
the fact that the this ``naive'' reduced \ps\ is constructed using the
functions $(p_\theta,f(\phi), {\rm P}^\pm_\phi)$; these are the
classical analogs of the Dirac observables constructed above, and as
we noted earlier, they fail to be a complete set. Recall that this
incompleteness occurs at points where $R(\phi)=0$. An analysis of the
constraint surface in the vicinity of these points shows that the two
``halves'' $\hat\Gamma_\pm$ of the reduced
\ps\ are smoothly joined at these points and the reduced \ps\
is in fact a cotangent bundle over $S^1$ \cite{thesis}%
\footnote{The reduced configuration space may consist of
 multiple copies of $S^1$, depending on the number of nodes of
 $R(\phi)$.}!
What is the relation between the two quantum theories?  As we have
seen, the Dirac quantum theory corresponds to the quantization of a
particle whose \ps\ is the cotangent bundle over two disconnected
intervals. This ``incorrect'' \ps\ can be thought of as the cotangent
bundle over $S^1$, {\em but with two diametrically opposite points
removed}. To illustrate the relation, we can construct a momentum
operator, which in the Dirac theory has a doubly degenerate spectrum
with even eigenvalues, whereas in the reduced space theory this
operator has a nondegenerate spectrum with all integer
eigenvalues. Thus the two quantum theories are inequivalent, and there
appear to be no obvious means to make them equivalent.

\mysection{Issue of time and deparametrization}

For {\it ordinary} constrained systems, such as gauge theories, the
Hamiltonian ---which generates dynamics--- is distinct from the
constraint functions and therefore does not vanish on the constraint
surface.  On the other hand, there are theories in which the vanishing
of the Hamiltonian is itself a first class constraint function.  We
will refer to such theories as {\em dynamically constrained systems}
since the generator of the dynamical trajectories is now constrained
to vanish.  General relativity in the spatially compact case is an
outstanding example of such systems.

In dynamically constrained systems, to begin with, the notions of
gauge and time evolution are entangled. To bring out the resulting
difficulties, let us first recall some features of ordinary
constrained systems. In such systems, solving the constraints
---either classically, by constructing the reduced \ps\ (or,
equivalently, a cross-section of the gauge orbits), or in quantum
theory, by constructing the physical states, an operator algebra of
observables and an inner product on these states--- is a purely {\em
kinematical} procedure. Conceptually, this construction is divorced
from the dynamical structure of the theory which is dictated by the
Hamiltonian. In the classical theory, the Hamiltonian can be projected
unambiguously to the reduced \ps, and all physically interesting
dynamics can be considered to occur there.  In the quantum theory, the
corresponding Hamiltonian operator generates (unitary) evolution on
the Hilbert space of physical states.

In contrast, for systems in which the Hamiltonian is constrained to
vanish, kinematical considerations are intimately linked with the
dynamical structure of the theory. (For a more complete discussion of
the problems in quantum theory, see \cite{kk:time,cji:time}.)  If one
proceeds as one does for ordinary constrained systems, one ends up
with a ``frozen formalism.'' Classically, each point in the reduced
phase space corresponds to an entire dynamical trajectory. Quantum
mechanically, solutions to the constraints can be found and represent
physical states, but they do not evolve. To obtain evolution, one must
re-interpret the constraint as telling us that how the ``true degrees
of freedom'' change with respect to an appropriate canonical variable
which can then be taken to represent time. Thus, for these systems,
``time'' is not an external parameter; it has to be singled out from
among the canonical variables.

In section 6.1, we will discuss the simplest of such systems in the
framework of algebraic quantization. We will see that one can follow
the program step by step and arrive at the inner product on the space
of physical states without having to single out time. In section 6.2,
we will discuss the issue of dynamics and interpretation. The choice
of our model was motivated by simplicity; we wish to illustrate the
ideas in as simple a setting as possible. For models which are
physically more interesting, see, e.g., \cite{atu:II}.

\subsection{Non-relativistic parametrized particle}
Consider a non-relativistic particle moving in a potential $V(q_i)$ in
Euclidean space. Dynamics is specified by the Hamiltonian $H({q_i},
{p_i}) = {\textstyle 1\over 2m}\Sigma{p_i p_i} + V({q_i})$. This
simple system can be ``parametrized'' by adding to the 3-dimensional
configuration space the time variable. Thus, the (enlarged)
configuration space, ${\cal C}$, is now 4-dimensional, coordinatized
by $(q_0, q_i)$; and the phase space is 8-dimensional. There is one
(first class) constraint:
 \be\label{npp:con}
       C(q,p):= p_0 + H(q_i,p_j) = 0,
 \ee
where $q$ and $p$ stand for $(q_0, q_i)$ and $(p_0, p_j)$
respectively. The constraint reduces the fictitious 4 degrees of
freedom to the original 3 ``true degrees'': classically, the
constrained system is equivalent to the original system evolving in
the 6 dimensional phase space spanned by $(q_i,p_i)$ via the
Hamiltonian $H(q_i, p_i)$.

Let us now carry out the quantization program step by step.  Let the
space \cS\ of elementary observables be the complex vector space
spanned by the 9 functions $(1,q,p)$ on the phase space $\Gamma$, with
the usual commutation relations. Choose for the representation space
\cV\ the space of smooth functions on the {\it 4-dimensional}
configuration space ${\cal C}$, and represent the operators by the
usual multiplication and partial derivative operators%
\footnote{Note that for what follows, in the choice of representation it is
 essential only that $\hq^0$ be a multiplication operator. One is free
 to choose any representation of the $\hq^i,\hp_i$ operators,
 depending on the specific form of the Hamiltonian.}.
The quantum constraint is now given by:
 \be \label{npp:qcon}
  \hC\circ\Psi (q)\equiv \dfrac\hbar{i}\> {\d\Psi (q)\over \d q^0} +
    \hH\circ \Psi(q)= 0\>.
 \ee
(Although (\ref{npp:qcon}) has the form of the Schr\"odinger equation
if $q^0$ is identified as the ``internal time,'' in this subsection we
will ignore this aspect and regard (\ref{npp:qcon}) simply as the
quantum constraint equation as in the previous examples.) The space of
physical states, ${\cal V}_{phy}$, now consists of solutions of this
equation. A technically convenient way to write them (formally) is:
 \be\label{npp:qsol}
  \Psi(q)=\eiq\circ\psi(q^i),
 \ee
where the $\psi(q^i)$ are arbitrary functions of $q^i$.  Note that the
solutions $\Psi(q)$ to the quantum constraint are complex valued
functions on the 4-dimensional configuration space ${\cal C}$; they
are {\it not} functions of $q_i$ alone as they necessarily depend on
$q^0$ as well. In this sense, they are ``covariant''. However, since
the $q^0$ dependence is fixed by the exponential term, physical states
are completely determined by the functions $\psi(q^i)$, which we can
think of as the ``initial data'' for the first order (in $q^0$)
differential equation (\ref{npp:qcon}).

Our next task in the quantization program is to isolate physical
observables. None of the elementary quantum operators, $\hq^0, \hq^i$
or $\hp_i$, corresponding to the set \cS, is a physical operator since
the action of any of them maps one out of the set of physical
states. Fortunately, it is not difficult to construct, at least
formally, a complete set of physical operators. They are given by:
\be \label{npp:pop1}
 \ba{rcl}
  \hQ^i(0)\circ\Psi & := & \hU(0)\hq^i \hU^{-1}(0)\circ\Psi  =\eiq
     \hq^i\circ\psi \equiv \eiq\circ q^i\psi(q^i)\mbox{ and}\\
  \hP_i(0)\circ\Psi & := & \hU(0)\hp_i \hU^{-1}(0)\circ\Psi  =\eiq
     \hp_i\circ\psi \equiv \eiq\circ \dfrac\hbar{i}\dfrac{\d}
  {\d q^i}\psi(q^i),
 \ea
\ee
where
 \be\label{npp:hatu}
  \hU(0):=\eiq.
 \ee
Here, since $\hq^0$ acts by multiplication, it has been replaced by
$q^0$ and the operator $\hat{H}$ is given by $\hH=H(\hq^i,\hp_i)$.
(The reason for the notation $\hU(0)$ will become clear in the next
subsection.) From the last step in (\ref{npp:pop1}), it is obvious that
$\hat{Q}^i(0)\circ \Psi$ and $\hat{P}_i(0)\circ\Psi$ are again
solutions to the quantum constraints; $\hat{Q}^i(0)$ and
$\hat{P}^i(0)$ are physical operators. Indeed, a simple algebraic
calculation shows that these six operators, $\hQ^i$ and $\hP_i$,
commute with the constraint, and furthermore, are their own
${}^\star$s. Since the reduced \ps\ is 6-dimensional, and the above
Dirac operators are independent, they form a complete set. Hence we
can now look for an inner product on $\cV_{phy}$ with respect to which
these operators are Hermitian. For this, let us begin by introducing a
measure $\mu(q)$ on the configuration space and set:
 \be \label{npp:ip1}
  \IP{\Psi (q)}{\Phi (q)} = \lint_{\cal C} d^4q\>
   \mu(q)\> \ovr\Psi (q)\> \Phi (q),
 \ee
for all physical states $\Psi (q)$ and $\Phi (q)$. To determine the
measure, we impose the Hermiticity requirements. The condition that
$\hQ^i$ be Hermitian does not constrain the inner product in any way.
The condition that $\hP_i$ be Hermitian requires that the measure be
independent of $q_i$. (In the general case, when the ``true''
configuration space is a non-trivial manifold or the coordinates are
not Cartesian, the Hermiticity conditions on $\hP_i$ determine the
dependence of $\mu$ on $q^i$. The important point is that the
dependence of $\mu$ on $q^0$ is left undetermined.) Thus, the
inner-product can now be calculated:
 \bea \label{npp:ip}
  \IP{\Psi (q)}{\Phi (q)} &=& {\disp \lint
  d^4q\> \mu (q^0)\> \ovr\Psi (q)\> \Phi (q)} \cr
  &=& {\disp \lint dq^0\>\mu (q^0) \lint d^3q^i \>\ovr\Psi (q^0,
     q^i)\> \Phi (q^0, q^i) } \cr
  &=& {\disp \lint dq^0\>\mu(q^0)  \lint d^3q^i
          \>\ovr\psi(q^i)\>\phi(q^i) } \cr
  &=& {\disp K \lint d^3q^i\>  \ovr\psi (q^i)\> \phi (q^i)
   \equiv K \lint d^3q^i\>  \ovr\Psi (q^0, q^i)\> \Phi (q^0, q^i) }
 \eea
where the constant $K$ is given by $K = \lint dq_0 \,\mu (q_0)$. Here,
in the third step, we have used the fact that $\Psi(q_0, q_i)$ and
$\Phi(q_0, q_i)$ are physical states, i.e., they satisfy
(\ref{npp:qcon}). Thus, the second integral in the second line is
independent of $q_0$. Since $\mu(q^0)$ is not constrained in any way
by the Hermiticity of the observables, we can choose it so that $K$ is
finite, say $K=1$. Thus, the reality conditions do indeed select a
unique inner product on $\cV_{phy}$ (up to the usual overall constant)
and the resulting quantum description is completely equivalent to the
quantum theory of the original unconstrained particle moving in a
potential $V$ in the Euclidean space.

The final picture is the following: the physical Hilbert space
consists of solutions to the constraint equation (\ref{npp:qcon}),
with the Hermitian inner product given by (\ref{npp:ip}). Up to this
point, the physical operators (\ref{npp:pop1}) were formal constructs,
used to find an inner product. Now, however, we can use the the
physical inner product to {\em rigorously} define the unitary operator
(\ref{npp:hatu}), and hence the physical observables. This completes
the quantization program.

We conclude this subsection with two remarks.

Note first that the Hermitian inner product is defined on \cVp, i.e.,
on the space of solutions $\Psi (q_0,q_i)$ to (\ref{npp:qcon}). That
in the final step we can perform the integral on a constant $q_0$
surface is ``accidental''; it is only a calculational device. The
situation is rather similar to that encountered in the covariant
symplectic description of fields on Minkowski space \cite{cov:sym}
where the expression of the symplectic structure involves an
integration over a spatial slice although the structure itself is
defined on the space of solutions to the field equations on the entire
space-time. In this sense, the above quantum description of the
parametrized particle is also ``covariant''.

In the main construction above, we have been dealing essentially with
the covariant states $\Psi(q)$.  Note, however, that these covariant
solutions are in 1-1 correspondence with the $q^0$ independent
(``initial data'') states $\psi(q^i)$. In fact, there is an obvious
unitary transformation, given by (\ref{npp:qsol}), between the
covariant states $\Psi(q)$ and the states $\psi(q^i)\equiv
\psi_0(q^i)$. The inverse of the unitary transformation is given by:
 \be \label{npp:ut0}
  \psi_0(q^i):=e^{\frac{i}{\hbar}\hH q^0}\circ\Psi(q)\equiv\left.
    \Psi(q)\right|_{q^0=0}.
 \ee
With $K=1$, the inner product on these states is simply
(\ref{npp:ip}). Clearly, the states $\psi_0(q^i)$ are not the
solutions of any constraint equation.  However, they carry a faithful
representation of the observable algebra. Let $z$ denote any operator
in the set $(q^i,p_i)$; and let $Z$ denote the corresponding operator
in the set $(Q^i,P_i)$.  Under the action of the unitary
transformation, the representation of the observables (\ref{npp:pop1})
is simply
 \be \label{npp:pop2}
  \hZ(0)\circ\psi_0(q^i) = \hz\circ\psi_0(q^i).  \ee
The physical observables have a simple action on the space of initial
states for the constraint equation.  Now, the intuitive meaning of
these operators is clear: Since the constraint generates dynamical
evolution, we know that the physical observables correspond to
constants of motion, which in turn can be identified with the position
and momentum at some initial time. Hence, a set of Dirac operators can
be obtained by ``evolving the covariant states $\Psi$ back to
$q^0=0$'' (or, via (\ref{npp:ut0}), evaluating them at $q^0=0$),
acting with the usual ``instantaneous'' operators on the initial state
$\psi(q^i)$, and then ``evolving the {\em resulting} initial state
forward to $q^0$'', using the constraint equation.  This is exactly
the procedure we have carried out, as is obvious also from the second
equalities in (\ref{npp:pop1}).

\subsection{From frozen formalism to dynamics}
We now wish to extract dynamics from the framework constructed above.
The main idea of course is to regard $q^0$ as the internal clock, $q^i$
as the ``true degrees of freedom,'' and interpret the constraint
equation as telling us how the functional dependence of the physical
states $\Psi$ on $q^i$ changes as we go from one $q^0 =
constant$ slice in the configuration space to another.

More precisely, to introduce the notion of evolution, it is necessary
to foliate the 4-dimensional covariant configuration space by
$q^0=constant$ surfaces. Then, each covariant state $\Psi(q)$ defines
a 1-parameter family of ``\Sch\ states'' $\psi_\tau(q^i)$:
 \be \label{npp:utt}
  \psi_\tau(q^i):=e^{\frac{i}{\hbar}\hH(q^0-\tau)}\circ\Psi(q)
        \equiv\left. \Psi(q)\right|_{q^0=\tau}.     \ee
Note that this correspondence exists {\em only} because the states
$\Psi(q)$ satisfy the constraint equation. Let us now turn to
observables.  If $\hat{z}$ is a ``Schr\"odinger observable,'' the
foliation naturally provides us with a 1-parameter family of physical
observables $\hat{Z}(\tau)$ via the inverse of the unitary
transformation (\ref{npp:utt}):
 \be \label{npp:cpop}
  \hZ(\tau)\circ\Psi(q) = \hU(\tau)\hz\hU^{-1}(\tau)\circ\Psi(q)
     =e^{-\frac{i}{\hbar}\hH(q^0-\tau)}\hz\circ\psi_\tau(q^i),
 \ee
where
 \be
  \hU(\tau)=e^{-\frac{i}{\hbar}\hH(q^0-\tau)}  \ee
is the unitary transformation defined in (\ref{npp:utt}). It is now
manifest from this analogy that we can identify $q^0$ with the time
in the quantum theory and that the $\hZ(\tau)$ are the ``evolving''
Heisenberg operators. In fact since the effect of the unitary
evolution is simply to evaluate the covariant state at $q^0=\tau$,
i.e.,
 \be
  \hU^{-1}(\tau)\circ\Psi(q):=\psi_\tau(q^i)\equiv\Psi(q^0=\tau,q^i) \ee
the above rule to calculate the $\hZ(\tau)$ operators states that
their action is the following: Evaluate the covariant physical states
$\Psi(q^0,q^i)$ {\em at} $q^0=\tau$, act with the corresponding $\hz$
operators, and then evolve forward to $q^0$ again. Thus, for
$\hq^0(\tau)$ the above rule yields
 \be
  \hq^0(\tau)\circ\Psi(q)=\hU(\tau)\circ\tau\psi_\tau(q^i)=\tau\cdot\Psi(q).
 \ee

Of course, if we desire, we can also evaluate the action of the
operators $\hZ(\tau)$ on a {\em fixed} \Sch\ Hilbert space, say
corresponding to $q^0=\tau_0$:
 \be \label{npp:cpopt} \hZ(\tau)\circ\psi_{\tau_0}(q^i)=\left[
     e ^{\frac{i}\hbar\hH (\tau-\tau_0)} \hz e^{-\frac{i}\hbar\hH(z)
       (\tau-\tau_0)} \right]\circ\psi_{\tau_0}(q^i). \ee
As expected, we have lost all reference to $q^0$, and have obtained
the complete deparametrization of the theory to the usual text-book
picture.

Finally, note that it is trivial to extend this discussion to allow
for a $q_0$-dependence in the expression of the Hamiltonian (by
appropriately time-ordering the $U(\tau)$) or to replace the Euclidean
space by an $n$-manifold.

We conclude this subsection with three remarks.

In retrospect we see that we could have worked always in the covariant
picture, with the $\tau$-dependent Heisenberg operators defined in
(\ref{npp:cpop}). However, we would then have lost both the
interpretation of $q^0$ as time as well as the motivation for the
introduction of the ``evolving'' observables. It is in order to see
the unfolding of the dynamics which is hidden in the frozen formalism
that we have to break the covariance of the space of solutions and
introduce on this space a ``foliation'' corresponding to time
evolution and the resulting sequence of \Sch\ states.

Recall that nowhere in the kinematical construction to find the inner
product was it necessary to treat $q^0$ in a special manner. We found
the inner product without explicitly eliminating the ``time'' $q^0$.
That is, contrary to what is commonly done in the literature, we did
{\it not} integrate only over the true degrees of freedom $q^i$.
Rather, we used the the reality conditions on the space $\cV_{phy}$ of
solutions to the constraint to obtain the inner product. This is an
important point, since it illustrates that it is not necessary to
isolate time in order to construct the Hilbert space of physical
states. In this simple example, the form of the constraint immediately
suggests that we treat $q^0$ as the internal clock and $q^i$ as the
true degree of freedom. Hence we could also have first singled out the
time variable and then found the inner product.  However, in more
interesting examples of dynamically constrained theories such as
3-dimensional general relativity, the constraints do not provide an
obvious internal clock and, except in the simplest spatial topology,
we do not yet know how to find a convenient deparametrization. One
can, however, impose the reality conditions directly and the strategy
does yield an unique inner product.  (For a more complete discussion
of this aspect of the issue of time in \qg, see \cite[\S
12]{aa:osgood,newbook}.)

However, to complete the analysis and make physical predictions, as in
the \Sch\ picture, one may need to find explicit solutions by
diagonalizing the ``true'' Hamiltonian $\hat{H}$. In addition to the
states, one has to construct explicit expressions for a complete set
of interesting operators. In Bianchi models \cite{atu:II} of
4-dimensional general relativity, for example, while the reality
conditions lead one directly to the inner product on the physical
Hilbert space, deparametrization is necessary to answer physically
interesting questions concerning the fate of classical singularities
in quantum theory. Furthermore, even from a mathematical
viewpoint, the availability of deparametrization simplifies the
quantization program significantly. In particular, it provides a
direct route to the problem of finding a complete set of physical
observables.

\mysection{Conclusion}

In this paper, we have illustrated various features of the algebraic
quantization program of \cite{newbook} through a number of examples.
The fact that the program could be carried out to completion in all
these examples provides confidence in the viability of the strategies
involved.

The main lessons of this study can be summarized as follows:
\begin{description}
\item[{\sl i)}] Overcompleteness of elementary classical variables can be
incorporated into quantum theory through appropriate algebraic
conditions on the elementary quantum operators. Thus, it is not
necessary to eliminate these relations classically. Indeed, in the
case when the phase space is a non-trivial manifold, it is in
principle impossible to do so. This point is conceptually important
for a number of systems being investigated in the literature,
including lattice gauge theories, where the Wilson loop functionals
form an overcomplete set on the configuration space and in continuum
Yang-Mills theories and general relativity, where the ``loop
variables'' of Gambini and Trias \cite{GT} and Rovelli and Smolin
\cite{RS} also form an over-complete set almost everywhere on the \ps.
\item[{\sl ii)}] Even when there are no obvious symmetries present, the
inner product on physical states can be singled out using the
``reality conditions,'' i.e., by demanding that real classical
observables be represented on the physical Hilbert space by
self-adjoint operators. In the constrained rotor model, in particular,
this procedure clarified an important conceptual point thereby
resolving a controversy.
\item[{\sl iii)}] The issue of completeness of physical observables
is subtle.  Even when the set of observables is ``locally complete,''
superselection sectors can arise due to global ambiguities. This issue
is important for general relativity where the Rovelli-Smolin loop
variables fail to constitute a complete set on sets of measure
zero \cite{GLS}.
\item[{\sl iv)}] Deparametrization of a dynamically constrained is not
essential to obtain a mathematically complete quantum description,
including the inner product on the space of physical states. However,
to display dynamics explicitly and extract physical information from
the theory, one may have to deparametrize the theory at least
approximately.  Furthermore, if an exact deparametrization happens to
be available, it can be {\em used} to find a complete set of (``time
dependent'') Dirac observables and these in turn can be used to obtain
the inner product on the space of physical states.
\end{description}

The program and the examples were motivated primarily by various
problems one encounters in quantization of general relativity. Some of
the points listed above have already played an important role in the
quantization of 3-dimensional general relativity \cite[\S17]{newbook},
and various mini-superspace models \cite{atu:II,gamm,gmt:b2,cr:prd}
of 4-dimensional gravity. We expect that these points are all
significant to the quantization of full, 4-dimensional general
relativity.

The program, however, has a much broader range of applicability; it is
not tied to general relativity. Indeed, since it is formulated rather
loosely, it serves more as an umbrella that brings together the ideas
underlying various approaches to the quantization of constrained systems
such as the group theoretic approach \cite{isham:lh} and geometric
quantization \cite{njmw,aa:ms}. Thus, the large class of
examples in which these methods have been successful also provide
illustrative applications of the program. However, because it is
formulated somewhat loosely, it allows more general strategies. For
example, unlike geometric quantization, it can be used to construct
``exotic'' representations that do not directly arise from
polarizations of the phase space. An important example is provided by
the loop representations of gauge theories \cite{GT}.

\mysection{Acknowledgments} We would like to thank Chris Isham, Karel
Kucha\v r, Jorma Louko, Joe Romano and Carlo Rovelli for
discussions. This work was supported in part by NSF grants PHY
93-96246, PHY 90-08502 and by the Eberly research fund of Penn State
University. RST would like to thank the members of the Center for
Gravitational Physics and Geometry for their hospitality and good cheer.

\end{document}